\documentclass[preprintnumbers]{revtex4}

\usepackage[pdftex]{graphicx}
\usepackage{epsf}
\usepackage{amsmath,amssymb}
\usepackage{bm}

\newcommand{\bvec}{\boldsymbol}

\begin{document}
\preprint{KUNS-2525}

\title{$2\alpha+t$ cluster feature of $3/2^-_3$ state in $^{11}$B}
\author{Yoshiko Kanada-En'yo}
\affiliation{Department of Physics, Kyoto University, Kyoto 606-8502, Japan}
\author{Tadahiro Suhara}
\affiliation{Matsue College of Technology, Matsue 690-8518, Japan}

\begin{abstract}
We reanalyze $2\alpha+t$ cluster features of $3/2^-$ states in $^{11}$B by 
investigating the $t$ cluster distribution around a $2\alpha$ core 
in $^{11}$B, calculated with the method of antisymmetrized molecular dynamics (AMD). 
In the $3/2^-_3$ state, a $t$ cluster is distributed in a wide region around $2\alpha$, 
indicating that the $t$ cluster moves rather freely in angular
as well as radial motion. 
From the weak angular correlation and radial extent of the $t$ cluster distribution, 
we propose an interpretation of a $2\alpha+t$ cluster gas for the $3/2^-_3$ state.
In this study, we compare the $2\alpha+t$ cluster feature in $^{11}$B($3/2^-_3$)
with the $3\alpha$ cluster
feature in $^{12}$C($0^+_2$), and discuss their similarities and differences.
\end{abstract}
\maketitle

\section{Introduction}
Many cluster structures have been found in light nuclei, such as 
$^7$Li with an $\alpha+t$ cluster and $^8$Be with a $2\alpha$ cluster.
In the excited states of $^{12}$C, various $3\alpha$ cluster structures have been discovered
(for example, Refs.~\cite{Fujiwara80,Freer:2014qoa} and references therein). 
In the early stages, the possibility of a linear 3$\alpha$ chain structure was proposed 
for $^{12}$C($0^+_2$) by Morinaga {\it et al.} \cite{morinaga56,morinaga66}. 
However, the linear chain structure of $^{12}$C($0^+_2$) has been excluded by 
the $\alpha$ decay width of this state \cite{suzuki72}. 
Later theoretical studies with $3\alpha$ cluster models have  
revealed that $^{12}$C($0^+_2$) is a weakly bound $3\alpha$ state 
with neither geometric structures of the linear chain nor triangle structures 
\cite{kamimura-RGM1,kamimura-RGM2,uegaki1,uegaki2,uegaki3,descouvemont87,kurokawa04,kurokawa05,Arai:2006bt}. 
Further extended models that do not rely on assumptions of the existence of clusters
have also obtained similar results of the cluster feature in $^{12}$C$(0^+_2)$
 \cite{Enyo-c12,KanadaEn'yo:2006ze,Neff-c12,Chernykh:2007zz}. 

In 2001, Tohsaki {\it et al.} proposed a new interpretation of $^{12}$C($0^+_2$) 
\cite{Tohsaki01}, i.e., the "$\alpha$ cluster gas" in which   
$\alpha$ clusters are weakly interacting like a gas
\cite{Funaki:2002fn,Yamada:2003cz,Funaki:2003af,Funaki:2009zz,Yamada:2011bi}.
In such a dilute cluster system, $\alpha$ particles behave as bosonic 
particles; therefore, this state has been discussed in relation to the $\alpha$ condensation predicted in dilute infinite matter \cite{Ropke98}. 
The cluster gas phenomenon has been attracting great interest because it is a new concept of the cluster state and is different from 
the traditional concept of geometric cluster structures, in which clusters are localized and have a specific spatial configuration.
Indeed, in this decade, theoretical and experimental 
studies of nuclei, such as $^8$He, $^{10}$Be, $^{11}$B, and $^{16}$O, have included 
intensive searching for cluster gas states  \cite{Kawabata:2005ta,KanadaEn'yo:2006bd,KanadaEn'yo:2007ie,Itagaki:2007rv,Wakasa:2007zza,Funaki:2008gb,Funaki:2010px,Yamada:2010vc,Ohkubo:2010zz,Suhara:2012zr,Ichikawa:2012mk,Kobayashi:2012di,Kobayashi:2013iwa}.

For $^{11}$B, a developed $2\alpha+t$ cluster structure in the $3/2^-_3$ state was predicted by a $2\alpha+t$ cluster model \cite{nishioka79} and the method of antisymmetrized molecular dynamics 
(AMD) \cite{KanadaEn'yo:2006bd,Suhara:2012zr}. 
The $3/2^-$ state at 8.56 MeV is assigned to this $2\alpha+t$ state because
the experimentally measured $M1$ and monopole transition strengths \cite{Kawabata:2005ta,Kawabata:2004pc} and 
the $GT$ transition strength of the mirror state \cite{Fujita04}
are reproduced by the calculation. Studies have found similarity between 
$^{11}$B($3/2^-_3$) and 
$^{12}$C($0^+_2$) in remarkable monopole transitions 
and also predicted a nongeometric cluster feature of
the $2\alpha+t$ structure in $^{11}$B($3/2^-_3$) similar to that of 
the $3\alpha$ cluster structure in $^{12}$C($0^+_2$)
\cite{Kawabata:2005ta,KanadaEn'yo:2006bd,Suhara:2012zr}.
From the analogies to $^{12}$C($0^+_2$), $^{11}$B($3/2^-_3$) was interpreted as
a $2\alpha+t$ cluster gas.

However, the cluster gas in a $2\alpha+t$ system 
has not been fully understood. 
The following problems remain to be clarified.
First, the parity of the $2\alpha+t$ state is negative and conflicts 
with the original idea of cluster gas, in which 
$\alpha$ clusters occupy an $S$ orbit \cite{Yamada:2010vc} and form a positive parity state. 
How can we extend the $\alpha$ cluster gas picture to the $3/2^-$ state containing a $t$
cluster with negative parity?
Second, it is not obvious whether the $2\alpha+t$ cluster state shows a nongeometric feature
because a state with three clusters is more bound in the $2\alpha+t$ system than in the $3\alpha$ system, resulting in 
a smaller size of $^{11}$B($3/2^-_3$) than that of $^{12}$C($0^+_2$)
because of the deeper effective potential between $t$ and $\alpha$ clusters 
than that between two $\alpha$ clusters.

In this study, we reanalyze the $^{11}$B and $^{12}$C wave functions obtained by AMD calculations in 
previous studies \cite{KanadaEn'yo:2006bd,KanadaEn'yo:2006ze} and investigate the motion of a $t$ cluster around $2\alpha$.
We pay particular attention to the angular motion of the $t$ cluster with respect to the $2\alpha$ orientation 
to judge whether the $2\alpha+t$ structures in $^{11}$B have a geometric feature with an angular correlation 
or a nongeometric structure with weak angular correlation similar to the $3\alpha$ gas state of $^{12}$C$(0^+_2$). We also discuss 
the analogy and differences between  $2\alpha+t$ structures in  $^{11}$B
and $3\alpha$ structures in $^{12}$C. 
By considering the body-fixed plane of three clusters, we can connect  
the $P$-wave motion of the $t$ cluster around $2\alpha$ to the $S$-wave motion of the $\alpha$ cluster around $2\alpha$, and then extend the picture of the $3\alpha$ cluster gas to the $2\alpha+t$ system. 

The paper is organized as follows. We describe the AMD and cluster models 
in Section \ref{sec:formulation}.
Section \ref{sec:results} discusses 
cluster structures of $3/2^-$ states of $^{11}$B, in comparison with those of $0^+$ states of $^{12}$C. 
The paper concludes with a summary in section \ref{sec:summary}.

\section{Formulation}\label{sec:formulation}
In Refs.~\cite{KanadaEn'yo:2006bd,KanadaEn'yo:2006ze}, 
$3/2^-$ states of $^{11}$B and $0^+$ states of $^{12}$C have been calculated 
with the AMD model.  
In the present study, we reanalyzed the AMD wave functions of $^{11}$B and $^{12}$C obtained in previous studies
by using the $2\alpha+t$ and $3\alpha$ cluster model wave functions written by the
Brink-Bloch (BB) model \cite{Brink66}.
In this section, we briefly explain the AMD model adopted in previous studies 
and describe the BB cluster wave function used in the present analysis.

\subsection{AMD model}
The AMD model is a useful approach to describe the formation and breaking of clusters
as well as shell-model states having non-cluster structures \cite{KanadaEnyo:1995tb,KanadaEnyo:1995ir}. 
The applicability of the AMD method to light nuclei has been proven \cite{ENYOsup,AMDrev,KanadaEn'yo:2012bj}.
In the previous studies of $^{11}$B and $^{12}$C, the variation after spin-parity projection (VAP) in the AMD framework
was applied \cite{KanadaEn'yo:2006bd,KanadaEn'yo:2006ze}. 
For the detailed formulation of the AMD+VAP, please refer to the previously mentioned references.

A Slater determinant
of Gaussian wave packets gives an AMD wave function of an $A$-nucleon system:
\begin{eqnarray}
 \Phi_{\rm AMD}({\bvec{Z}}) &=& \frac{1}{\sqrt{A!}} {\cal{A}} \{
  \varphi_1,\varphi_2,...,\varphi_A \},\\
 \varphi_i&=& \phi_{{\bvec{X}}_i}\sigma_i\tau_i,\\
 \phi_{{\bvec{X}}_i}({\bvec{r}}_j) & = &  \left(\frac{2\nu}{\pi}\right)^{4/3}
\exp\bigl\{-\nu({\bvec{r}}_j-\frac{{\bvec{X}}_i}{\sqrt{\nu}})^2\bigr\},
\label{eq:spatial}\\
 \sigma_i &=& (\frac{1}{2}+\xi_i)\sigma_{\uparrow}
 + (\frac{1}{2}-\xi_i)\sigma_{\downarrow},
\end{eqnarray}
where $\phi_{{\bvec{X}}_i}$ and $\sigma_i$ are spatial and spin functions
of the $i$th single-particle wave function, respectively, and 
$\tau_i$ is the isospin
function fixed to be up (proton) or down (neutron). 
Accordingly, an AMD wave function
is expressed by a set of variational parameters, ${\bvec{Z}}\equiv 
\{{\bvec{X}}_1,{\bvec{X}}_2,\ldots, {\bvec{X}}_A,\xi_1,\xi_2,\ldots,\xi_A \}$.
The width parameter $\nu$ is chosen to be a common value for all nucleons and it is 
taken to be $\nu=0.19$ fm$^{-2}$ for $^{11}$B and $^{12}$C.

In the AMD+VAP method, we perform
energy variation after spin-parity projections in the AMD model space
to obtain the wave function for the lowest $J^\pi$ state.
Namely, the parameters ${\bvec{X}}_i$ and $\xi_{i}$ ($i=1,\ldots,A$) of the AMD wave function 
are varied to minimize the energy expectation value,
$\langle \Phi|H|\Phi\rangle/\langle \Phi|\Phi\rangle$,
with respect to the spin-parity eigenwave function projected 
from an AMD wave function, $\Phi=P^{J\pi}_{MK}\Phi_{\rm AMD}({\bvec{Z}})$.
For excited  $J^\pi_k$ states, the variation is performed for the energy expectation value of 
the component of the projected AMD wave function 
$P^{J\pi}_{MK}\Phi_{\rm AMD}({\bvec{Z}})$,
orthogonal to the lower $J^\pi_i$ ($i=1,\ldots,k-1$) states.

In each nucleus, all AMD wave functions obtained for various $J^\pi$ states are 
superposed to obtain final wave functions $\Psi_{\rm AMD+VAP}(J^\pi_k)$ 
for $J^\pi_k$ states by solving the Hill-Wheeler equation. To describe $^{11}$B and $^{12}$C, approximately, twenty independent AMD wave functions are adopted for basis wave functions 
in the superposition \cite{KanadaEn'yo:2006bd,KanadaEn'yo:2006ze}.
Because the number of basis AMD wave functions is finite, 
continuum states cannot be treated properly in the present AMD+VAP method.
The method is a bound state approximation in which resonance states are  
obtained as bound states. 

In the AMD model space, we treat all single-nucleon wave functions as 
independent Gaussian wave packets, and therefore, cluster formation and breaking 
are described by spatial configurations of Gaussian centers, $\bvec{X}_i$. 
If we choose a specific set of parameters $\{\bvec{Z}\}$, 
the AMD wave function can be equivalent to a BB cluster wave function.

Note that the cluster-breaking component affects not only the ground state but also  
the excited cluster states of $^{12}$C, as shown in Refs.~\cite{Chernykh:2007zz,Suhara-new}.
We should stress that the AMD model can describe various 
cluster structures while incorporating cluster-breaking effects.

\subsection{Cluster model}

To analyze $t$-cluster motion in the $^{11}$B wave functions obtained  
by the AMD+VAP model, 
we measure the position of the $t$ cluster around the $2\alpha$ core
by using the BB wave function, which expresses
a three-center cluster wave function with a specific spatial configuration of 
cluster positions. 

The BB wave function for a $2\alpha+t$ cluster structure of $^{11}$B is described as 
\begin{equation}\label{eq:BB}
|\Phi_{\rm BB}(\bvec{R}_1,\bvec{R}_2,\bvec{R}_3)\rangle =\frac{1}{\sqrt{A!}}{\cal A}
\left\{\psi_\alpha(\bvec{R}_1)\psi_\alpha(\bvec{R}_2)\psi_t(\bvec{R}_3) \right\}\rangle.
\end{equation}
Here $\psi_\alpha(\bvec{R}_i)$ and $\psi_t(\bvec{R}_i)$ are 
$\alpha$- and $t$-cluster wave functions described by the harmonic oscillator  $(0s)^4$ and $(0s)^3$ 
shell-model wave functions with the shifted center position $\bvec{R}_i$, respectively.
The width parameter of the harmonic oscillator is $\nu=0.19$ fm$^{-2}$, same as the 
AMD wave functions for $^{11}$B and $^{12}$C in the previous studies.

When clusters are far from each other and the antisymmetrization effect between clusters is negligible,  
$\bvec{R}_i$ indicates cluster center positions around which the $\alpha$ and $t$ clusters are localized.
In other words, the parameters $\bvec{R}_i$ $(i=1,2,3)$ specify the spatial configuration of cluster positions. 
Note that if $\bvec{R}_i$ are close to each other, the antisymmetrization effect is strong
and $\bvec{R}_i$ does not necessarily have a physical meaning of the cluster position. 
For instance, in the small distance limit between cluster centers $\bvec{R}_i$, 
the BB wave function no longer describes localized clusters but rather 
it describes a shell-model wave function. 

\begin{figure}[tb]
\begin{center}
	\includegraphics[width=6.5cm]{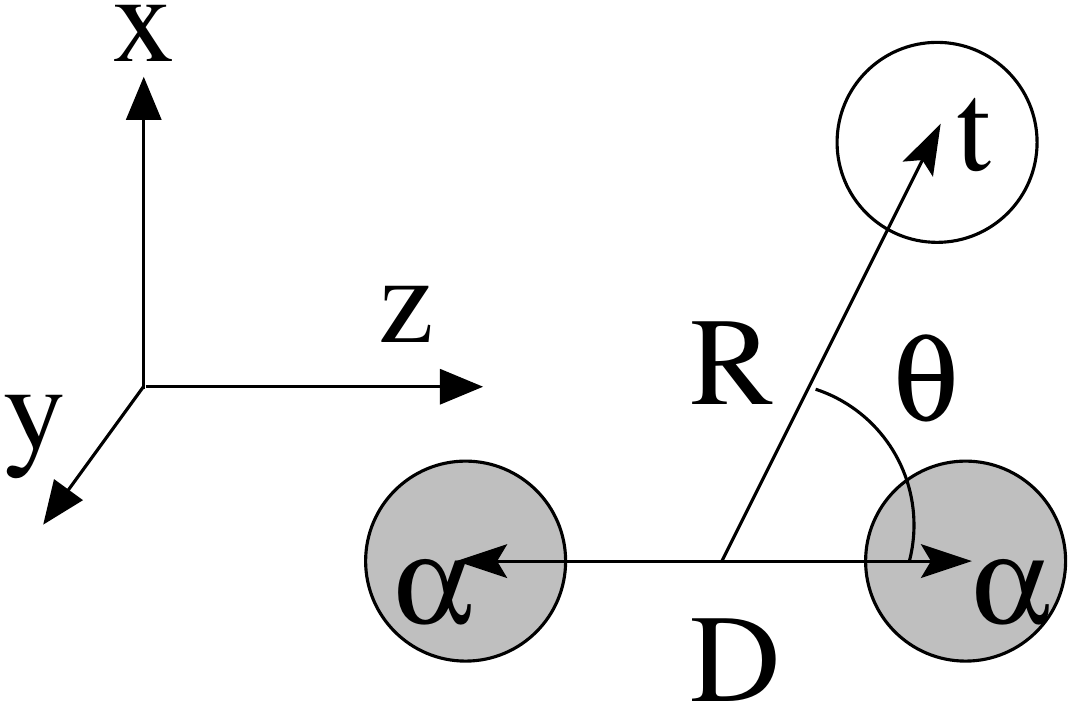} 	
\end{center}
\vspace{0.5cm}
  \caption{Schematic figure of $2\alpha+t$ cluster structure.
\label{fig:2a-t}}
\end{figure}

We consider a configuration of three clusters, $2\alpha$ and $t$, on the $z\text{--}x$ plane at $y=0$ 
with $(4\bvec{R}_1+4\bvec{R}_2+3\bvec{R}_3)/11=0$.
Two $\alpha$ clusters are set at the distance $D$ 
with the orientation parallel to the $z$-axis as
$\bvec{R}_1-\bvec{R}_2=(0,0,D)$, and the $t$ cluster is located  
at the distance $R$ from the center of the $2\alpha$ (see Fig.~\ref{fig:2a-t}).
We define the $t$-cluster position $\bvec{R}$ relative to the $2\alpha$ center as 
\begin{equation}
\bvec{R}\equiv \bvec{R}_3-\frac{\bvec{R}_1+\bvec{R}_2}{2}=(R\sin \theta, 0, R\cos \theta),
\end{equation}
where the $t$-cluster direction is chosen at an angle $\theta$  
from the $z$-axis (the $2\alpha$ orientation).

To measure $t$ cluster probability at a certain position, 
we calculate the overlap of the wave function,
$\Psi_{\rm AMD+VAP}(3/2^-_k)$, for $^{11}$B($3/2^-_k$)
with the $J^\pi$-projected BB wave function
\begin{equation}\label{eq:overlap}
U(D,\bvec R)=\langle P^{J\pi}_{MK}\Phi_{\rm BB}(\bvec{R}_1,\bvec{R}_2,\bvec{R}_3)|\Psi_{\rm AMD+VAP}(J^\pi_k)\rangle,
\end{equation}
for $J^\pi=3/2^-$. In the present study, the $t$ intrinsic spin is set to the
$+z$ direction and $K=+3/2$ is chosen so as to fix the $z$-component of the 
$t$ orbital angular momentum to be $L^{(t)}_z=+1$.

If a $t$ cluster is located far from $2\alpha$ and the antisymmetrization effect between $t$ and $2\alpha$ 
is negligible, the relative wave function between the $t$ cluster and $2\alpha$ 
in the BB wave function is given by a Gaussian 
$\exp[-\gamma(\bvec{r}-\bvec{R})^2]$ with $\gamma=\sqrt{24/11}\nu$, i.e., 
the $t$ cluster is well localized around $\bvec{R}$. 
Therefore, the BB wave function can be regarded as the 'test function' for the $t$ cluster, located at 
$\bvec R$ from the $2\alpha$
core with the $\alpha$-$\alpha$ distance $D$. The overlap of the test function 
with $\Psi_{\rm AMD+VAP}(3/2^-_k)$ is roughly regarded as the $t$ wave function 
on the $z\text{--}x$ plane at $y=0$, and its square stands for the $t$-cluster probability at $\bvec R$ around the $2\alpha$ core.

In the present analysis, the normalization of $\Phi_{\rm BB}$ is determined by 
the normalization of the constituent cluster wave functions 
$\psi_\alpha(\bvec{R}_i)$ and $\psi_t(\bvec{R}_i)$ before the antisymmetrization and projections.
Namely, norms of $\psi_\alpha(\bvec{R}_i)$ and $\psi_t(\bvec{R}_i)$ are chosen independently of 
$\bvec{R}_i$ and are kept to be constant. They are determined 
to make the norm of the BB wave function 
$\Phi_{\rm BB}(\bvec{R}_1,\bvec{R}_2,\bvec{R}_3)$ to be a unit at the limit
where cluster positions are far from each other and the antisymmetrization effect vanishes.

As cluster positions become close to each other, the norm of the BB wave function becomes
smaller because of the antisymmetrization effect, which means that 
the cluster wave function, i.e., the $t$-cluster probability, 
is suppressed by Pauli blocking effect between nucleons in different clusters.

For a $3\alpha$ system, we analyze the motion of an $\alpha$ cluster around the $2\alpha$ core 
in a similar way to the $t$ cluster in the $2\alpha+t$ system. 
The BB wave function for a $3\alpha$ system is given by 
replacing the $t$ cluster in Eq.~\ref{eq:BB} with an $\alpha$ cluster having a position
$\bvec{R}_3$ of $(4\bvec{R}_1+4\bvec{R}_2+4\bvec{R}_3)/12=0$.
A configuration of three $\alpha$ clusters on the $z\text{--}x$ plane at $y=0$ 
is considered, and we define the distance parameter $D$ 
for the $2\alpha$ core and
the position $\bvec{R}$ for the $\alpha$ cluster around the $2\alpha$  in the
same way as the $2\alpha+t$ case.
To measure the $\alpha$ cluster probability at a certain position around the $2\alpha$ core, we calculate
the overlap of the wave function 
$\Psi_{\rm AMD+VAP}(0^+_k)$ for $^{12}$C($0^+_k$)
with the $0^+$-projected BB wave function. 
The overlap is regarded as the $\alpha$-cluster wave function 
on the $z\text{--}x$ ($y=0$) plane, and its square stands for the probability 
of an $\alpha$ cluster at $\bvec{R}$ moving around the $2\alpha$ core.

\section{Results}\label{sec:results}

\subsection{AMD+VAP wave functions for $^{11}$B($3/2^-$) and $^{12}$C($0^+$)}

We analyzed the AMD+VAP wave functions  $\Psi_{\rm AMD+VAP}(J^\pi_k)$ 
for $3/2^-_1$,  $3/2^-_2$, and $3/2^-_3$ states of $^{11}$B, and $0^+_1$, $0^+_2$, 
and $0^+_3$ states of 
$^{12}$C, which were obtained in previous studies
\cite{KanadaEn'yo:2006bd,KanadaEn'yo:2006ze}.

The adopted effective nuclear interactions is 
the MV1 force (case 3) \cite{MVOLKOV} 
of the central force, supplemented by the spin-orbit term of the G3RS force \cite{LS}.
The interaction parameters used in Ref.~\cite{KanadaEn'yo:2006bd} for $^{11}$B are $m=0.62$, $b=h=0.25$, and $u_{\rm ls}=2800$ MeV, 
and those in Ref.~\cite{KanadaEn'yo:2006ze} for $^{12}$C are $m=0.62$, $b=h=0$, and $u_{\rm ls}=3000$ MeV. 
We slightly modified the parameter set for $^{11}$B from those for $^{12}$C to obtain a better reproduction  
of energy levels of $^{11}$B, as described in Ref.~\cite{KanadaEn'yo:2006ze}. 

The $^{11}$B wave functions are given by the 
superposition of $J^\pi=3/2^-$ eigenwave functions projected from 17 basis AMD wave functions 
obtained by the VAP calculations, whereas
$^{12}$C wave functions are described by 
$J^\pi=0^+$ eigenwave functions projected from 23 basis AMD wave functions.

\begin{figure}[tb]
\begin{center}
	\includegraphics[width=6.5cm]{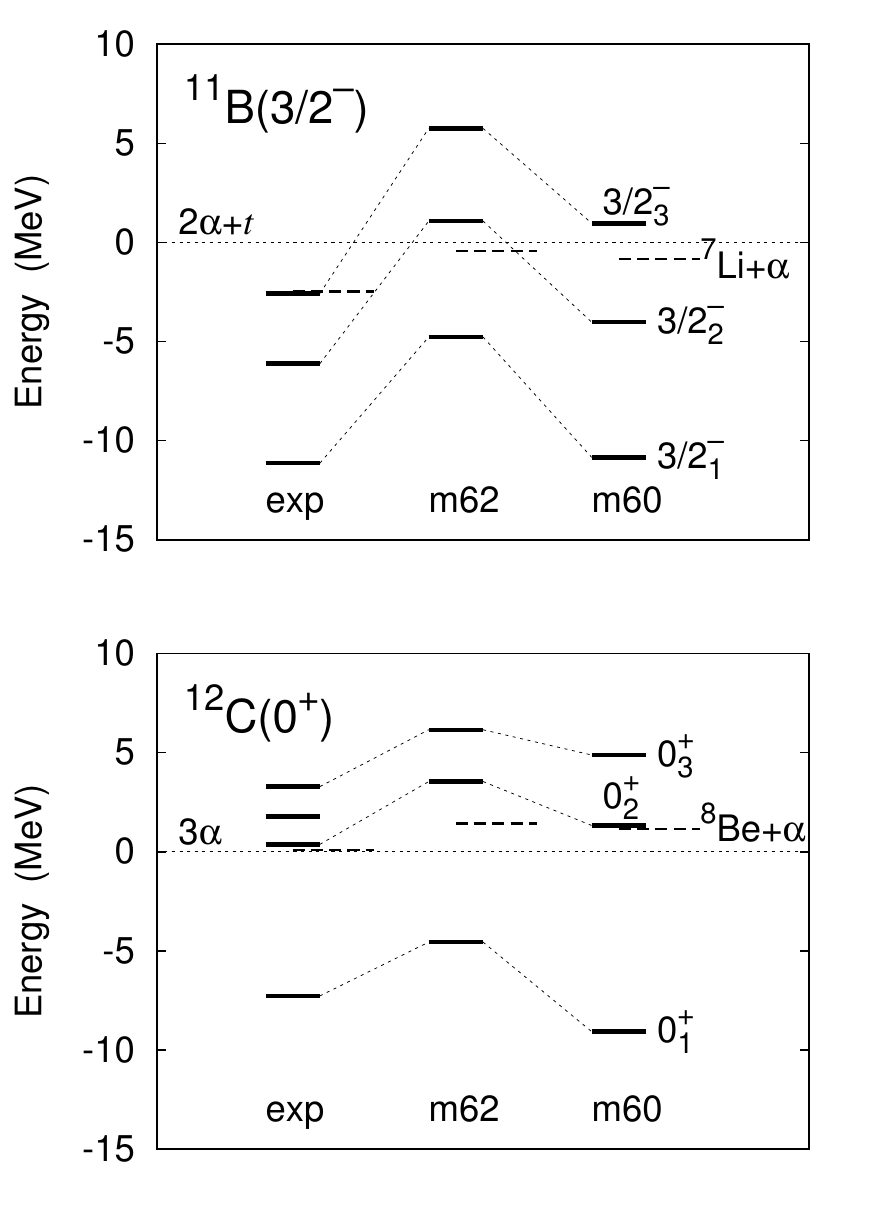} 	
\end{center}
\vspace{0.5cm}
  \caption{Energies of $3/2^-$ states of $^{11}$B measured from the $2\alpha+t$ threshold 
and those of $0^+$ states of $^{12}$C measured from the $3\alpha$ threshold.
$^7$Li+$\alpha$ and $^8$Be+$\alpha$ threshold energies are also shown. 
Theoretical values calculated using $m=0.62$ and those for $m=0.60$ are shown
in comparison with the experimental energy levels.
\label{fig:spe}}
\end{figure}

Figure ~\ref{fig:spe} shows the energy levels for $3/2^-$ states of $^{11}$B
and $0^+$ states of $^{12}$C. 
Compared with the experimental data, 
the interaction used in the previous studies tends to overestimate the relative energies to the threshold energies. 
We can improve the overestimation by changing the Majorana parameter
$m$ of the MV1 force. 
To obtain deeper binding 
wave functions than the original $m=0.62$ results,
we also perform calculations of $^{11}$B and $^{12}$C 
with the modified value $m=0.60$
in solving the Hill-Wheeler equation, using the same basis AMD wave functions 
as in the original studies. 
We call the original parametrization "m62" and the modified one "m60". 
Figure ~\ref{fig:spe} also shows the energy levels obtained with the m60 interaction. 
Interaction modification improves the energy positions of excited states.
Because of the deeper binding, sizes of excited states and
the radial motion of the $t$-cluster are slightly shrunk in the m60 result, as shown later.
However, the feature of 
the $t$-cluster motion around $2\alpha$ in $^{11}$B 
is qualitatively unchanged by the modification from m62 to m60; therefore, in the present study, we mainly discuss the original m62 result.

Table \ref{tab:radii} shows the calculated root mean square (rms) radii of 
matter density and rms charge radii. The experimental charge radii for the ground states are also listed. 
$^{11}$B($3/2^-_3$), $^{12}$C($0^+_2$), and  $^{12}$C($0^+_3$) have remarkably larger radii 
than the ground states because these states have developed cluster structures.

As discussed in Ref.~\cite{KanadaEn'yo:2006ze}, 
 $^{12}$C($0^+_2$) has no geometric structure, but it is 
described by the superposition of various configurations of three $\alpha$ clusters. 
This is consistent with calculations of the $3\alpha$ cluster model \cite{uegaki1} and the fermionic molecular dynamics
\cite{Chernykh:2007zz}
and the cluster gas picture proposed by Tohsaki {\it et al.} \cite{Tohsaki01}.
In contrast to $^{12}$C($0^+_2$), for $^{12}$C($0^+_3$), the AMD calculation predicts
a geometric cluster feature, having a large overlap with the $3\alpha$ 
cluster wave function with a chain-like $3\alpha$ configuration.

For $^{11}$B($3/2^-_3$), the AMD+VAP wave function shows 
a geometric feature quite similar to 
those of $^{12}$C($0^+_2$). This wave function is described by the
superposition of various configurations of 2$\alpha$ and $t$ clusters, which
means that
three clusters are weakly interacting like a gas. 
Another analogy of $^{11}$B($3/2^-_3$) to $^{12}$C($0^+_2$) is strong monopole transition 
from the ground state \cite{Kawabata:2005ta}. This transition 
means that $^{11}$B($3/2^-_3$) is 
understood by radial excitation, similarly to $^{12}$C($0^+_2$), rather than 
by angular excitation. 
Because of these analogies of $^{11}$B($3/2^-_3$) to $^{12}$C($0^+_2$),
an interpretation of the $2\alpha+t$ cluster 
gas state for $^{11}$B($3/2^-_3$) was proposed in the previous studies.

\begin{table}[ht]
\caption{The rms radii of matter density and rms charge radii
calculated with the AMD+VAP using m62 and m60 interactions.
The experimental charge radii are taken from Ref.~\cite{nucl-data-charge-radius}.
The unit is fm.
\label{tab:radii}}
\begin{center}
\begin{tabular}{cccccc}
\hline
\hline
 &  \multicolumn{2}{c}{m62} &  \multicolumn{2}{c}{m60} & exp. \\
  & matter & charge & matter & charge & charge \\
 $^{11}$B$(3/2^-_1)$ & 2.5 & 2.6 & 2.5 & 2.6 & 2.406 (0.291)  \\
 $^{11}$B$(3/2^-_2)$ & 2.7 & 2.8 & 2.7 & 2.8 & $-$ \\
 $^{11}$B$(3/2^-_3)$ & 3.0 & 3.1 & 3.0 & 3.0 & $-$ \\

 $^{12}$C$(0^+_1)$ & 2.5 & 2.7 & 2.5 & 2.6 & 2.47 (0.022)  \\
 $^{12}$C$(0^+_2)$ & 3.3 & 3.4 & 3.0 & 3.2 & $-$ \\
 $^{12}$C$(0^+_3)$ & 4.0 & 4.1 & 3.9 & 4.0 & $-$ \\
\hline
\end{tabular}
\end{center}
\end{table}

\subsection{$\alpha+\alpha$ and $\alpha+t$ cluster systems}

Because $\alpha$-$t$ and $\alpha$-$\alpha$ effective interactions may give 
essential contributions to three-body cluster dynamics of $2\alpha+t$ and $3\alpha$ systems,  
we here describe properties of subsystems, $\alpha+t$ and $2\alpha$ systems, obtained by 
the present nuclear interactions.

$^7$Li and $^8$Be are 
described well by $\alpha+t$ and $2\alpha$ cluster models 
and they are regarded as weakly (quasi) bound 
two-body cluster states. The ground states of $^7$Li and $^8$Be are
the $3/2^-_1$ and $0^+_1$ states, respectively. 
The experimental $^7$Li energy measured from the 
$\alpha+t$ threshold is $-$2.47 MeV and the $^8$Be energy from the $2\alpha$ threshold is 0.093 MeV.
These facts indicate that 
clusters are bound relatively deeper in the $\alpha+t$ system than 
in the $2\alpha$ system.

We calculate $^7$Li$(3/2^-_1)$ and $^8$Be($0^+_1$) with
the generator coordinate method (GCM) using $\alpha+t$ and $2\alpha$ cluster wave functions given by the BB model.
The width parameter $\nu=0.19$ fm$^{-2}$, same as for $^{11}$B and $^{12}$C wave functions,
is used for $\alpha$ and $t$ clusters. The distance parameters $D_{\alpha\text{-}\alpha}$ 
and $D_{\alpha\text{-}t}$ for the generator coordinate 
is taken to be $D_{\alpha\text{-}\alpha(t)}$=1, 2, $\ldots$, 8 fm, and resonance states are obtained as bound states within a bound state approximation in a finite volume. 
In the GCM calculation, the $t$-$\alpha$ binding obtained in $^7$Li is deeper
than the $\alpha$-$\alpha$ binding in $^8$Be. 
The energies of $^7$Li and $^8$Be measured from the 
$\alpha$-decay threshold are $-$0.4 MeV and 1.4 MeV for the m62 case and 
$-$0.8 MeV and 1.2 MeV for the m60 case.

\begin{figure}[tb]
\begin{center}
	\includegraphics[width=6.5cm]{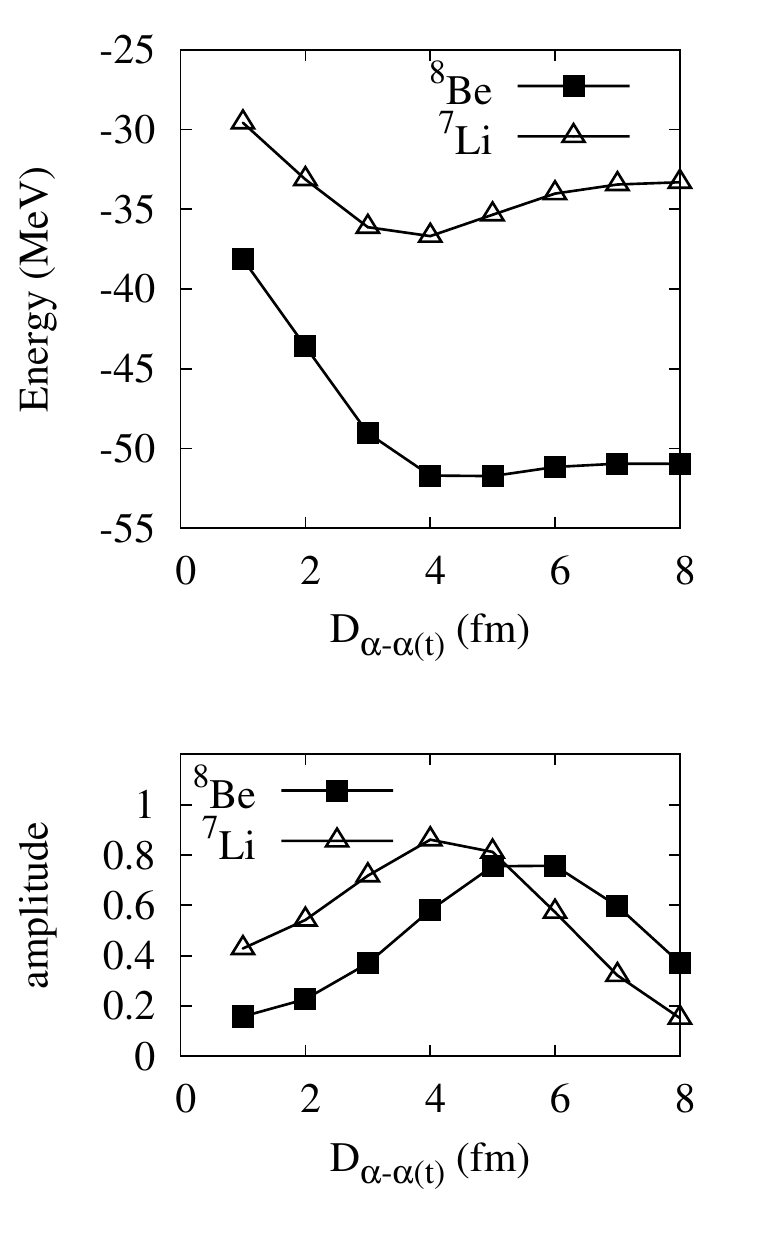} 	
\end{center}
\vspace{0.5cm}
  \caption{(Upper) energy curve of the $\alpha+t$ system for $^7$Li
and that of the $2\alpha$ system for $^8$Be.
The $J^\pi=3/2^-$ and $J^\pi=0^+$ projected energies are plotted as
functions of the intercluster distance $D_{\alpha\text{-}\alpha(t)}$ of the BB cluster wave function. 
(Lower) GCM amplitudes for the $^7$Li($3/2^-$) state were obtained by the 
 $\alpha+t$ cluster GCM calculation, and those for the 
 $^8$Be($0^+$) state were obtained by the $2\alpha$ cluster GCM calculation.
\label{fig:li7-be8}}
\end{figure}

Figure~\ref{fig:li7-be8} shows the $J^\pi=3/2^-$ energy projected from a single 
BB model wave function for $\alpha+t$ and the $0^+$ energy for $2\alpha$ 
as functions of the intercluster distance $D_{\alpha\text{-}\alpha(t)}$. The figure also shows
GCM amplitudes defined by the squared overlap of a $J^\pi$-projected BB wave function 
with the GCM wave functions.
Here the $J^\pi$-projected BB wave function is normalized to have a unit norm.
The energy curve shows the effective repulsion in a small distance region
because of the Pauli blocking effect from the antisymmetrization.
The effective repulsion is larger in the $2\alpha$ system because of the stronger Pauli blocking 
and it pushes clusters outward, as seen in the GCM amplitudes. 
Also in the $\alpha+t$ system, clusters are developed spatially and 
distributed in an outer region because of the 
antisymmetrization effect in the inner region. Quantitatively,  the $\alpha+t$ system is
relatively deeper bound than the $2\alpha$ system because of the
weaker Pauli blocking as well as the attractive spin-orbit force. 

\subsection{Cluster motion in $^{11}$B and $^{12}$C}

In the previous study, we interpreted $^{11}$B($3/2^-_3$) as the $2\alpha+t$ cluster gas state
because of the analogies of $^{11}$B($3/2^-_3$) to $^{12}$C($0^+_2$) in cluster features and 
the monopole transition. 
One characteristic of these cluster states is that the nongeometric 
feature of three cluster configuration is different from $^{12}$C($0^+_3$), which 
has a large overlap with the open triangle configuration.
If the $2\alpha+t$ system has a nongeometric cluster structure, the $t$ cluster around the 
$2\alpha$ has weak angular correlation, and it shows wide distribution 
in angular motion. Therefore, the $t$ distribution in the nongeometric cluster structure 
should be different from that in a geometric cluster structure,
which concentrates at a certain angle,
reflecting a specific configuration of cluster positions.
In other words, the wide distribution of clusters in angular motion 
is an evidence of the nongeometric cluster structure, and it can be 
a probe for a cluster gas state, which is characterized by 
weak correlations in angular motion as well as in radial motion.

In the present analysis of cluster motion in $2\alpha+t$ and $3\alpha$ 
systems, we consider the body-fixed frame in which the configuration of cluster positions
is parametrized by $\bvec{R}$ for the $t$ ($\alpha$) position and $D$ for 
the $\alpha$-$\alpha$ distance of the $2\alpha$ core, 
as explained before (see Fig.~\ref{fig:2a-t}).
In the body-fixed frame, the angular correlation is reflected by
the angular distribution of the $t$ ($\alpha$) cluster around the $2\alpha$ core.

\subsubsection{Energy surface}
\begin{figure}[tb]
\begin{center}
	\includegraphics[width=15cm]{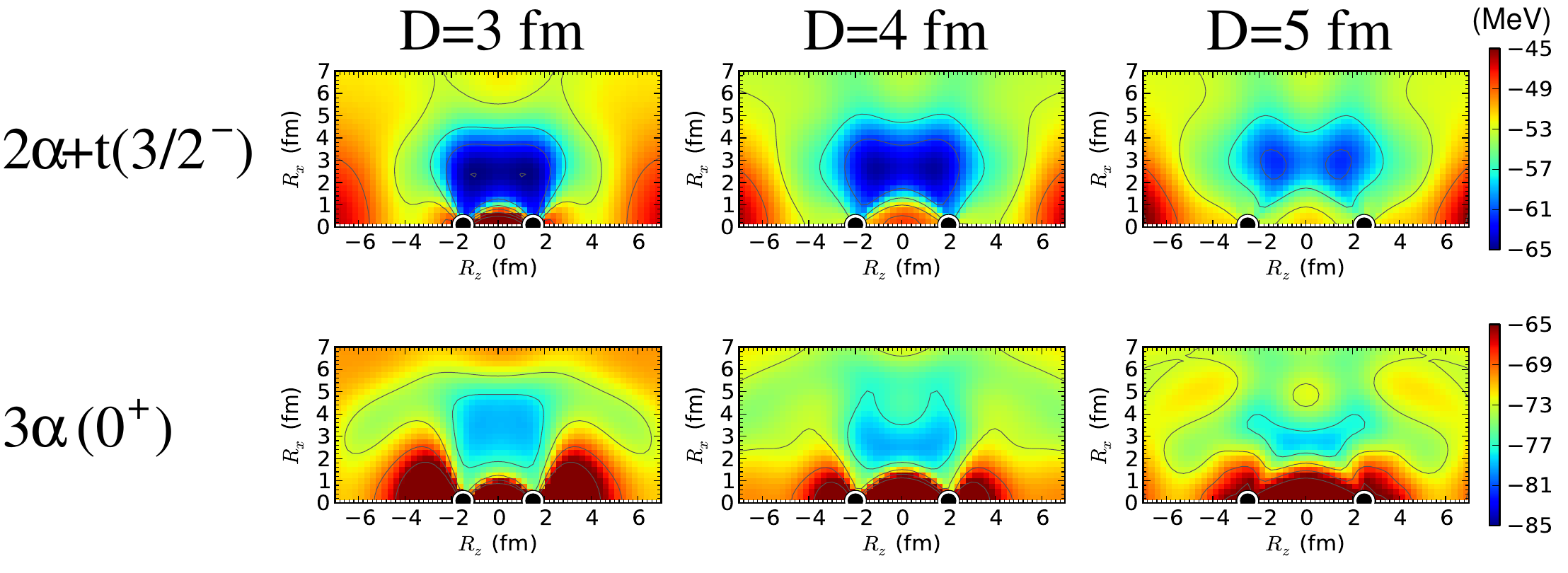} 	
\end{center}
\vspace{0.5cm}
  \caption{(Color online) Energy of the $(J^\pi,K)=(3/2^-,+3/2)$ states 
projected from BB wave functions for the $2\alpha+t$ cluster and that of 
the $0^+$ states for $3\alpha$ cluster calculated with m62. This figure shows
the energy surface on the $(R_z,R_x)$ plane for a fixed $2\alpha$ core
with the $2\alpha$ distance $D=3$, 4, and 5 fm.
\label{fig:ene}}
\end{figure}

Figure~\ref{fig:ene} shows the energy surface of the $2\alpha+t$ wave function 
on the ($R_z,R_x$) plane for the $t$ position around 
a fixed $2\alpha$ core with the $\alpha$-$\alpha$ distance 
$D$  and that of the $3\alpha$ wave function for the $\alpha$ motion around a $2\alpha$.
The energy of the $J^\pi$-projected BB wave function given as
\begin{equation}
E_{2\alpha+t(\alpha)}(D,\bvec{R})\equiv  \frac{
\langle \Phi_{\rm BB}(\bvec{R}_1,\bvec{R}_2,\bvec{R}_3) P^{J^\pi}_{MK}
|H| P^{J^\pi}_{MK}\Phi_{\rm BB}(\bvec{R}_1,\bvec{R}_2,\bvec{R}_3)\rangle 
}{\langle \Phi_{\rm BB}(\bvec{R}_1,\bvec{R}_2,\bvec{R}_3) P^{J^\pi}_{MK}
| P^{J^\pi}_{MK}\Phi_{\rm BB}(\bvec{R}_1,\bvec{R}_2,\bvec{R}_3)\rangle }
\end{equation}
for $(J^\pi,K)=(3/2^-,+3/2)$ states of $2\alpha+t$ and 
$J^\pi=0^+$ states of $3\alpha$ are calculated with the m62 interaction. 

In the energy surface of $2\alpha+t$ for the $t$ motion, 
the energy pocket at $R=2\text{--}3$ fm for the $2\alpha$ core with $D=3$ fm corresponds 
to the ground state of $^{11}$B. 
For the $2\alpha$ core with $D=4\text{--}6$ fm, 
the energy surface is soft in the $R=4\text{--}6$ fm region against the angular motion 
as well as the radial motion of the $t$ cluster. 
As shown later in this paper, $^{11}$B($3/2^-_3$) 
contains significant components of $2\alpha+t$ in this soft region.
In the $\theta \sim 0$ and $180^\circ$ region along the $z$-axis, the 
energy is relatively high,
indicating that the linear configuration is unfavored. 
Also, in the energy surface of $3\alpha$ for the $\alpha$ motion around
the $2\alpha$ core, the energy surface 
is soft for the $2\alpha$ core with $D=4\text{--}6$ fm in the angular and radial motions,
except for the $\theta\sim 0$ and $180^\circ$ regions.
It is shown that the $\theta\sim 0$ and $180^\circ$ configurations for the linear
structure are unfavored also in the $3\alpha$ system. 
From the softness of the energy surface, we expect that 
a $t$ cluster or an $\alpha$ cluster 
can move around the $2\alpha$ rather freely in the large $R$ region,
except for the $\theta\sim 0$ and $180^\circ$ regions.

It should be noted that the $(J^\pi,K)=(3/2^-,+3/2)$ state of $2\alpha+t$ 
has a node at the $z$-axis and its component vanishes 
in the $\theta=0$ and $180^\circ$ configuration for the 
ideal linear structure.
Although the energy of the $(J^\pi,K)=(3/2^-,+3/2)$ state can be calculated for 
the small $R_x$ limit, the parameter $R_x$ does not have 
a physical meaning of the $t$-cluster position in the $R_x<1/\sqrt{2\gamma}\sim 1$ fm region.

\subsubsection{Cluster motion in $^{11}$B and $^{12}$C}

To investigate cluster motion in $^{11}$B and $^{12}$C, we 
calculate the overlap $U(D,\bvec{R})$ 
between the BB wave function and the AMD+VAP wave function defined in 
Eq.~\ref{eq:overlap}.
We focus on the $\bvec{R}$ dependence of $U(D,\bvec{R})$ 
to see the $t$-cluster motion in $^{11}$B and to 
compare it with that of the $\alpha$-cluster motion in $^{12}$C.

\begin{figure}[tb]
\begin{center}
	\includegraphics[width=15cm]{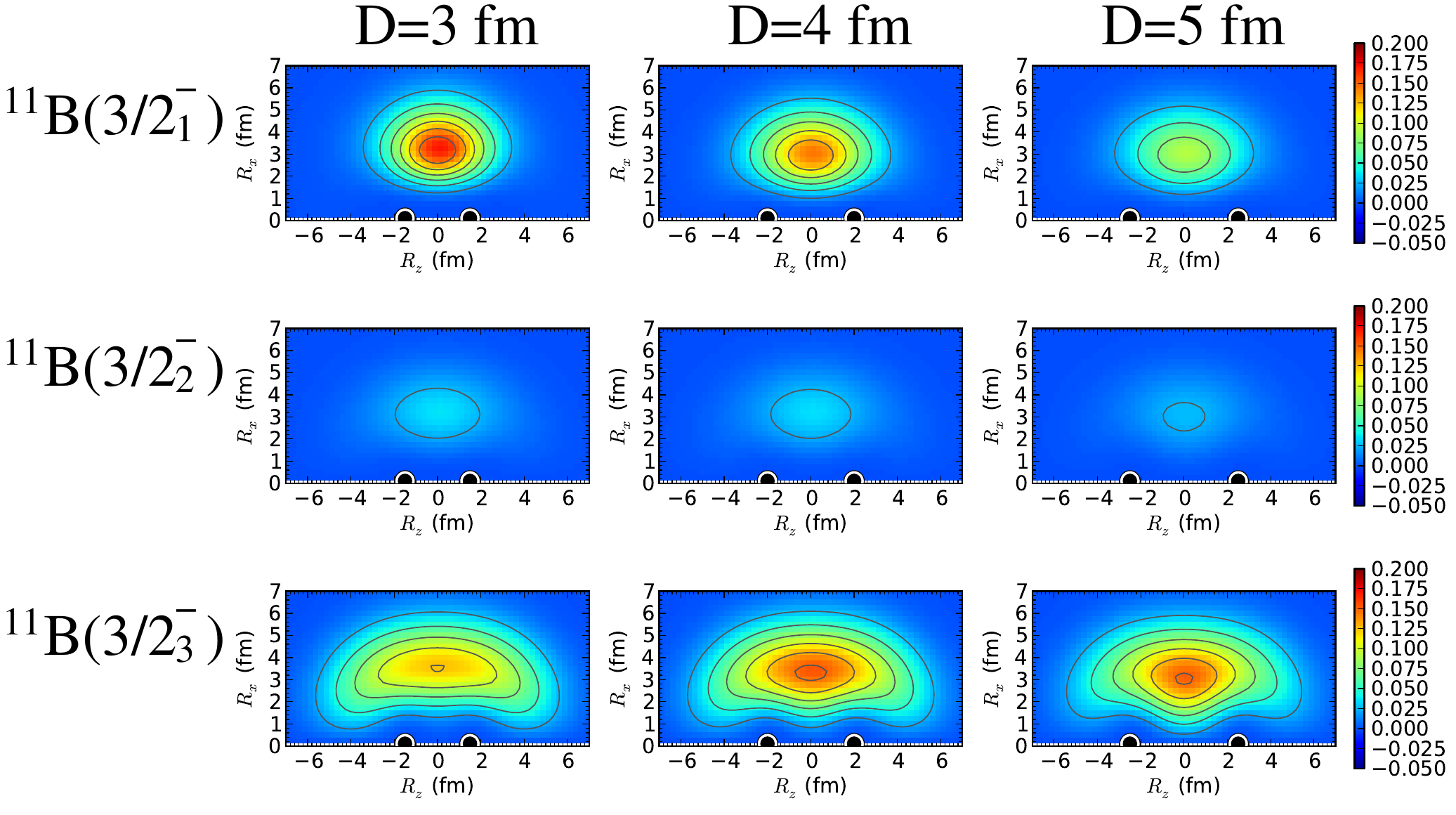} 	
\end{center}
\vspace{0.5cm}
  \caption{(Color online) Overlap $U(D,\bvec{R})$ of the BB wave function for 
 $2\alpha+t$ cluster with the AMD+VAP wave functions for $^{11}$B($3/2^-_1$),  
 $^{11}$B($3/2^-_2$), and $^{11}$B($3/2^-_3$) calculated with the m62 interaction.
 The overlap is plotted on the $(R_z,R_x)$ plane for $D=3$, 4, and 5 fm. 
 The $\alpha$ positions $(R_z,R_x)=(\pm D/2,0)$ in the $2\alpha$ core are shown by 
black circles.
\label{fig:b11-over}}
\end{figure}

\begin{figure}[tb]
\begin{center}
	\includegraphics[width=15cm]{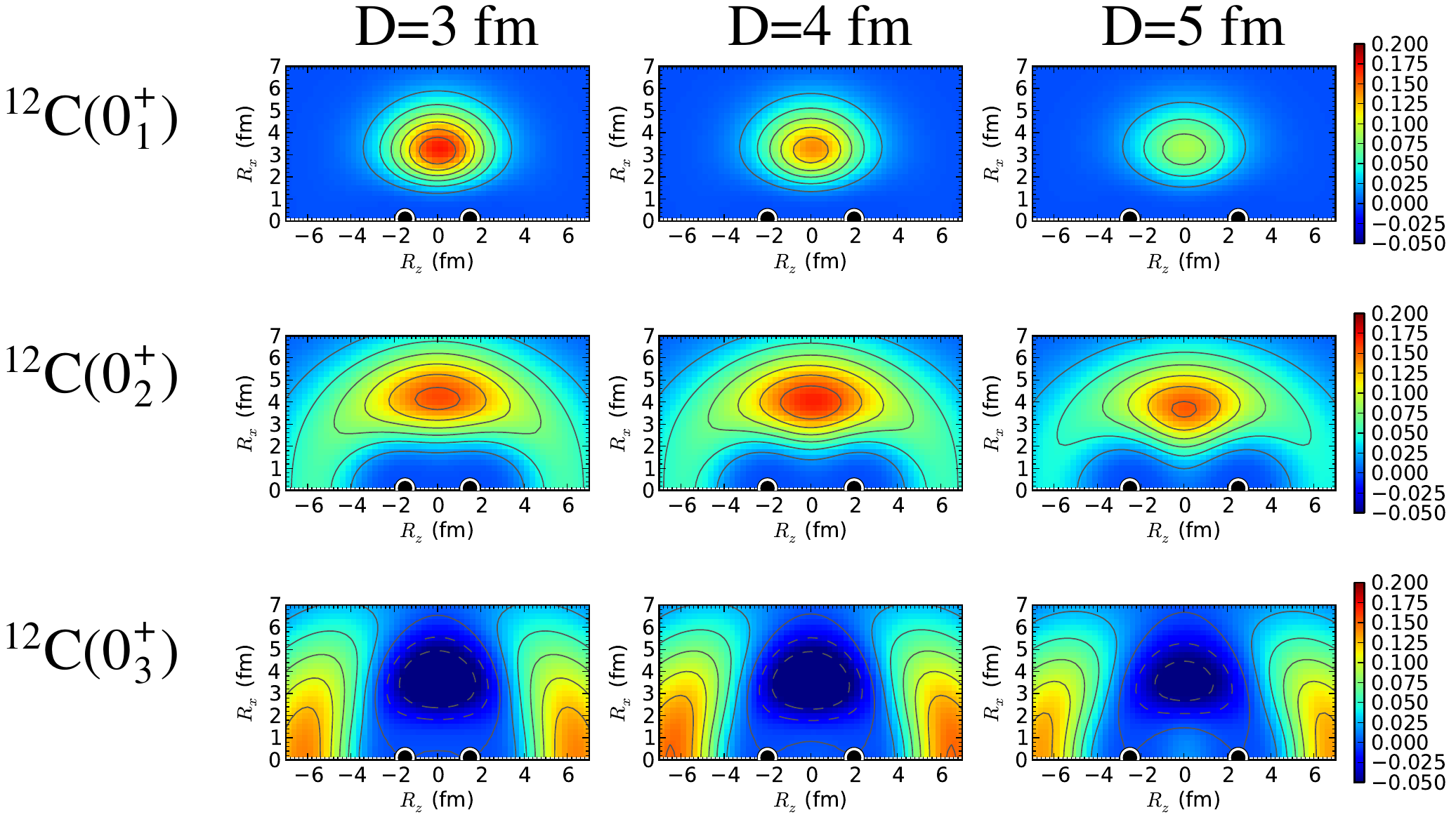} 	
\end{center}
\vspace{0.5cm}
  \caption{(Color online) Overlap $U(D,\bvec{R})$ of the BB wave function for a
 $3\alpha$ cluster with the AMD+VAP wave functions for $^{12}$C($0^+_1$),  
 $^{12}$C($0^+_2$), and $^{12}$C($0^+_3$), calculated with the m62 interaction.
 The overlap is plotted on the $(R_z,R_x)$ plane for $D=3$, 4, and 5 fm.
  The $\alpha$ positions $(R_z,R_x)=(\pm D/2,0)$ in the $2\alpha$ core are shown by 
black circles.
\label{fig:c12-over}}
\end{figure}

\begin{figure}[tb]
\begin{center}	
	\includegraphics[width=8.5cm]{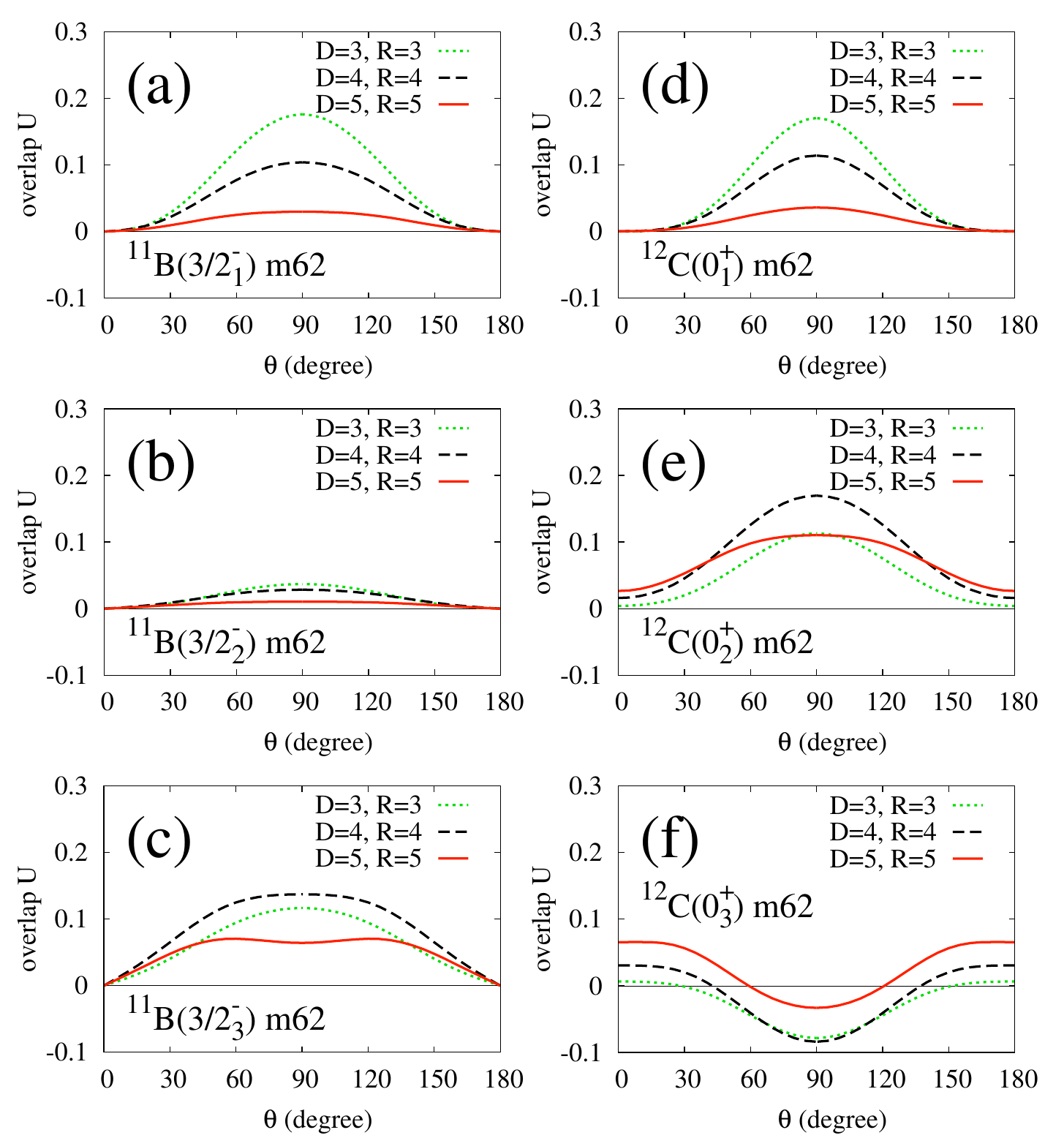} 	
\end{center}
\vspace{0.5cm}
  \caption{(Color online) $\theta$ dependence of overlap $U(D,\bvec{R})$ for $^{11}$B and $^{12}$C
 calculated with the m62 interaction
in the $R=D$ case.
\label{fig:angle}}
\end{figure}

We show the $R_x\text{--}R_z$ plot of the overlap $U(D,\bvec{R})$ for each $D$ 
for $^{11}$B($3/2^-_1$), $^{11}$B($3/2^-_1$), and $^{11}$B($3/2^-_3$) 
in Fig.~\ref{fig:b11-over} 
and that for $^{12}$C($0^+_1$),
$^{12}$C($0^+_2$), and $^{12}$C($0^+_3$)    
in Fig.~\ref{fig:c12-over}. 
The figures 
show the $t$-cluster motion and the $\alpha$-cluster motion 
around a $2\alpha$ core with the fixed $\alpha$-$\alpha$ distance 
$D$. We also show the $\theta$ dependence of the $U(D,\bvec{R})$ for 
$D=R=3$, 4, and 5 fm cases in Fig.~\ref{fig:angle}, which 
shows the angular distribution of the $t$ cluster in  $^{11}$B and the 
$\alpha$ cluster in $^{12}$C.

In $^{11}$B($3/2^-_1$), a remarkable peak at $(R_z,R_x) \sim (0,3)$ fm
for $D=3$ fm indicates that clusters are confined in the inner region to form a compact triangle 
configuration of $2\alpha+t$. 
In contrast, $^{11}$B($3/2^-_2$) has a small overlap with the $K=+3/2$ component of 
$2\alpha+t$ cluster wave functions because this state is dominated by the 
$L=2$ excitation of the $2\alpha$ core and has a large overlap with the $K=-1/2$ component
rather than with $K=+3/2$. This result is consistent with the discussion in previous study that 
$^{11}$B($3/2^-_2$) is the angular $\Delta L=2$ excitation, having the weak 
monopole transition from the ground state, and it has a 
radius as small as the ground state; therefore, it is not a cluster gas state.
In $^{11}$B($3/2^-_3$), the $t$ cluster is not localized 
and its component is distributed in a wide area of $R$ and $\theta$ 
in the $R=3\text{--}5$ fm region. The component in the inner region near the $2\alpha$ core
is suppressed because of the Pauli blocking between $t$ and $\alpha$ clusters.
Moreover, the component completely vanishes at the $R_x=0$ line because of the trivial nodal 
structure at the $z$-axis of the $t$ motion in the $L^{(t)}_z=+1$ orbit.
Nevertheless, compared with the ground state, 
the $^{11}$B($3/2^-_3$) wave function 
has a wide distribution of the $t$ cluster on the ($R_x,R_z$) plane, which 
means that the $t$ cluster is not localized, but it 
moves rather freely, excluding the $\theta\sim 0$ and $180^\circ$ regions. 
Furthermore, it contains significant $2\alpha+t$ components for
a large $\alpha$-$\alpha$ distance $D\ge 4$ fm of the $2\alpha$ core. 
This information indicates that two $\alpha$ clusters in the $2\alpha$ core are 
bound more weakly in $^{11}$B($3/2^-_3$) than in $^{11}$B($3/2^-_1$).

Features of $\alpha$-cluster distributions in $^{12}$C($0^+_1$) and $^{12}$C($0^+_2$) 
are similar to the $t$ distributions in $^{11}$B($3/2^-_1$) and $^{11}$B($3/2^-_3$), respectively.
In $^{12}$C($0^+_1$),  
a significant peak of the overlap $U(D,\bvec{R})$ exists at $(R_z,R_x)\sim (0,3)$ fm 
for $D=3$ fm, indicating the large overlap with a compact triangle 
configuration of $3\alpha$. 
In contract to the ground state, the $\alpha$ cluster 
is not localized,
but it is distributed in a wide area of $R$ and $\theta$ in $^{12}$C($0^+_2$). 
The component in the region close to the $z$-axis, i.e., 
in the $\theta\sim 0$ and $180^\circ$ regions, 
is suppressed because of the Pauli blocking effect from other $\alpha$ clusters in the core
and because the linear $3\alpha$ structure is energetically unfavored. As a result,
the $\alpha$ distribution in the $R\le 4$ fm region shows angular motion that is 
similar to that of the $t$ distribution in $^{11}$B($3/2^-_3$).
One characteristic of the $\alpha$ distribution in $^{12}$C($0^+_2$) that is 
different from the $t$ distribution in $^{11}$B($3/2^-_3$) is the significant 
$\alpha$ distribution in the $R\ge 6$ fm region. 
In this region, far from the $2\alpha$ core, 
the angular distribution of the $\alpha$ cluster 
in $^{12}$C($0^+_2$) becomes isotropic, showing $S$-wave feature
differently than that of $^{11}$B($3/2^-_3$).

With $^{12}$C($0^+_3$), the $\alpha$ distribution is 
concentrated in the $|R_z|=6\text{--}7$ fm
and $R_x \le 2$ fm region.  The distribution indicates a remarkably
developed $3\alpha$ cluster structure, however, the angular motion of the $\alpha$ cluster
in $^{12}$C($0^+_3$) is quite different from that in $^{12}$C($0^+_2$). 
$^{12}$C($0^+_3$) shows a strong angular correlation corresponding to the geometric 
configuration of the chain-like structure. This result contrasts with   
the weak angular correlation in $^{12}$C($0^+_2$), in which  
the $\alpha$ cluster is distributed in the wide $\theta$ region. 

Note that the remarkable peak for the compact triangle configuration in
the ground states of $^{11}$B and $^{12}$C 
originates in the antisymmetrization effect between clusters. In the region $R<3$ fm, 
the cluster wave function is almost equivalent to the shell-model wave function. 
Because the quantum effect is significant in this region, $\bvec{R}$ 
has less meaning of the localization or position of clusters in the classical picture. 

\begin{figure}[tb]
\begin{center}
	\includegraphics[width=8.5cm]{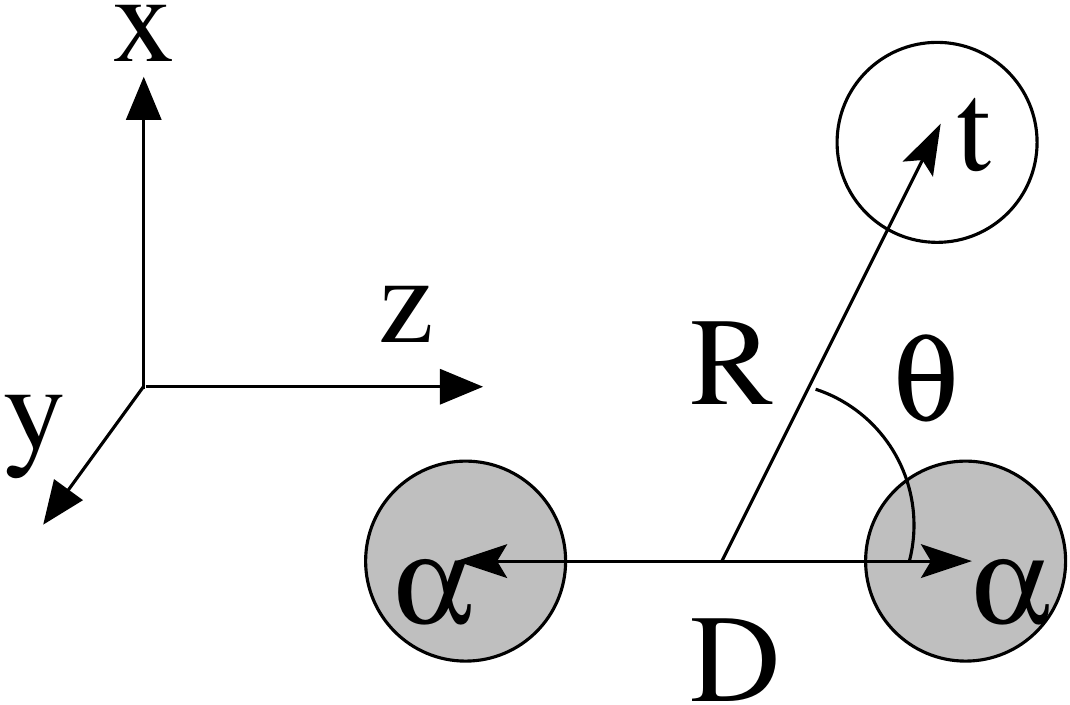} 	
\end{center}
\vspace{0.5cm}
  \caption{(Color online) Same as Fig.~\ref{fig:angle}; However, it is for the m60 result.
\label{fig:angle-m60}}
\end{figure}

We also performed the same analysis for the $^{11}$B and $^{12}$C wave functions 
obtained with the m60 interaction, and found that the cluster motion
is qualitatively the same as the m62 case, as shown in Fig.~\ref{fig:angle-m60} for the m60 result, which corresponds to Fig.~\ref{fig:angle} for the m62 result.

\subsubsection{Angular and radial motion in $^{11}$B($3/2^-_3$) and $^{12}$C($0^+_2$) }
\begin{figure}[tb]
\begin{center}
	\includegraphics[width=8.5cm]{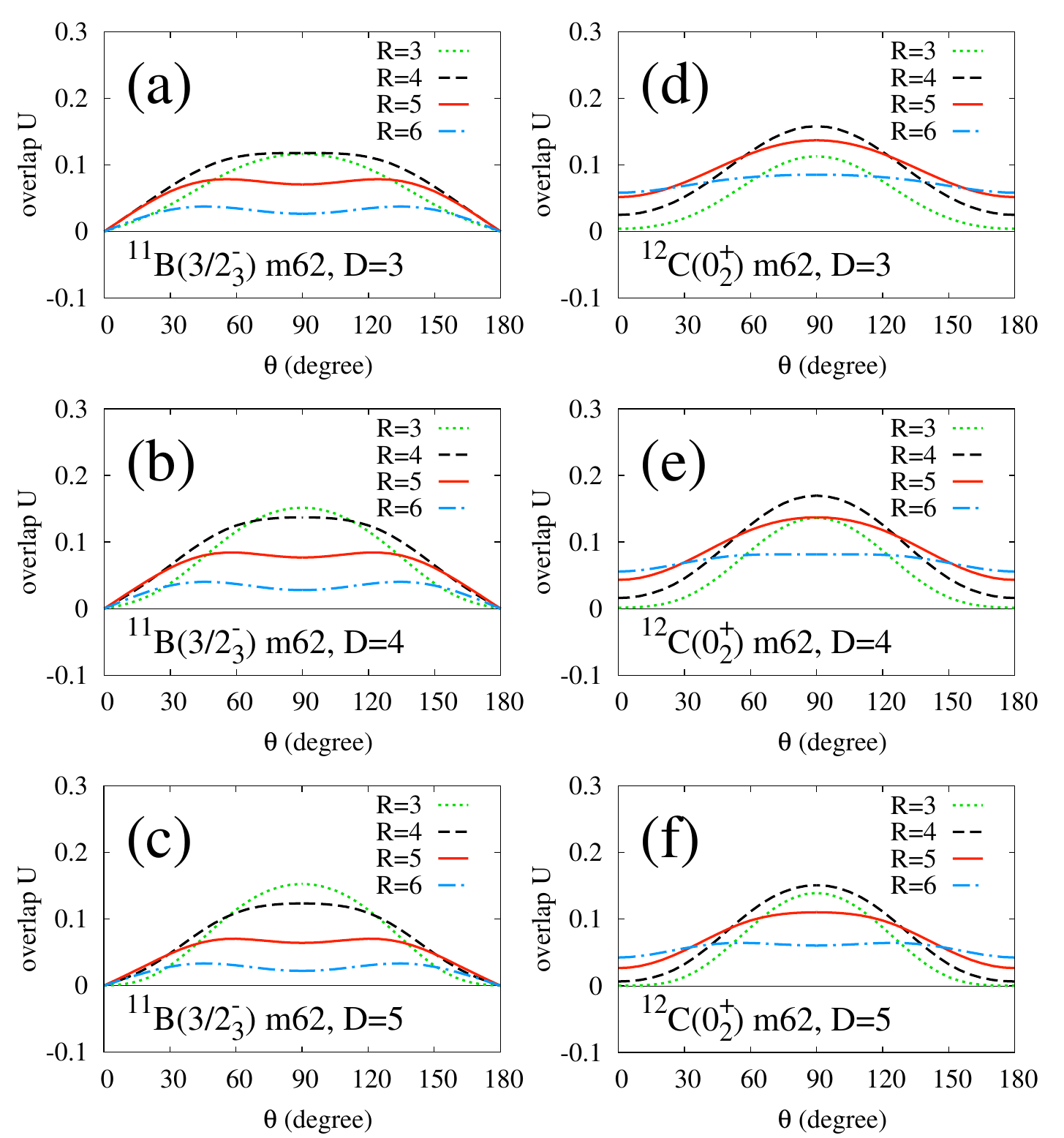} 	
\end{center}
\vspace{0.5cm}
  \caption{(Color online) $\theta$ dependence of overlap $U(D,\bvec{R})$ for $^{11}$B($3/2^-_3$) 
and $^{12}$C($0^+_2$)  calculated with the m62 interaction.
\label{fig:angle0}}
\end{figure}

\begin{figure}[tb]
\begin{center}
	\includegraphics[width=8.5cm]{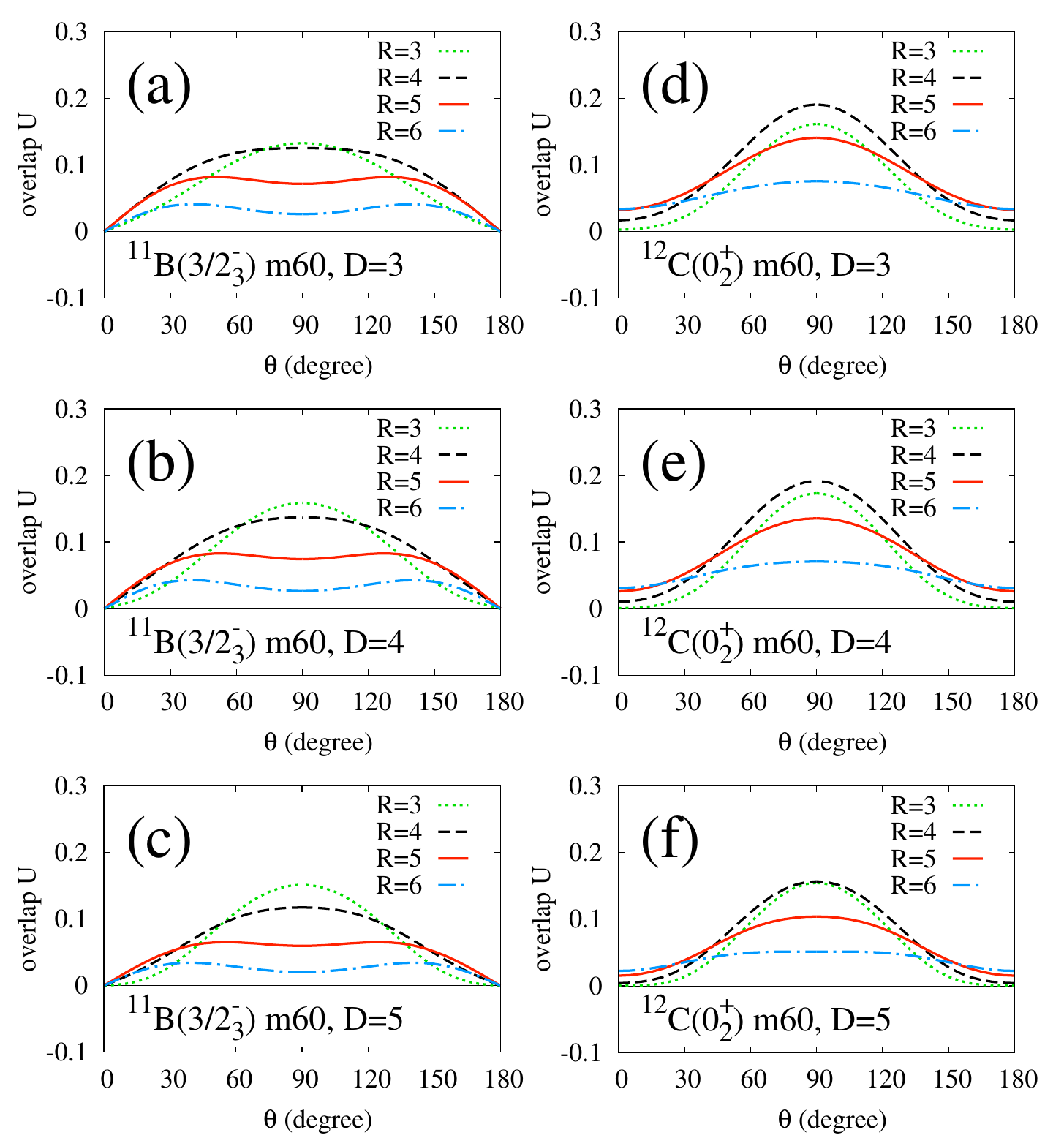} 	
\end{center}
\vspace{0.5cm}
  \caption{(Color online) Same as Fig.~\ref{fig:angle0} but for the m60 result.
\label{fig:angle0-m60}}
\end{figure}

To discuss the angular motion of the $t$ cluster 
in $^{11}$B($3/2^-_3$) and compare it 
with the $\alpha$ cluster in $^{12}$C($0^+_2$) in more detail, 
we show the $\theta$ dependence of the overlap $U(D,\bvec{R})$ for $R=3\text{--}6$ fm 
in each $D$ case in Fig.~\ref{fig:angle0} for m62 and Fig.~\ref{fig:angle0-m60} for m60.

At $R=3$ fm, the $t$ distribution in $^{11}$B($3/2^-_3$) 
has a peak structure at $\theta=90^\circ$, which means that the angular motion of the $t$ cluster 
is restricted in the narrow $\theta$ region 
because of the Pauli blocking effect from $\alpha$ clusters. 
At $R=4$ fm and 5 fm, the $t$ cluster is distributed widely. 
In particular, at $R\ge 5$ fm, the $t$ distribution 
becomes almost flat in the wide region of $30^\circ < \theta < 150^\circ$.
Because of the node structure of the $t$ motion in the $L^{(t)}_z=+1$ orbit, 
the $t$ distribution vanishes at $\theta=0$ and $180^\circ$ 
in the $(J^\pi,K)=(3/2^-,+3/2)$ component of the $2\alpha+t$ system. 
As a result, the $t$ cluster in $R\ge 5$ fm
moves almost freely in the wide $\theta$ region,
except for $\theta=0$ and $180^\circ$ regions for the trivial nodes.

In $^{12}$C($0^+_2$), the angular distribution of the $\alpha$ cluster at $R=3$ fm 
has a peak at $\theta=90^\circ$, similar to the $t$ distribution in $^{11}$B($3/2^-_3$). 
As $R$ increases, the angular distribution becomes wide. However, 
because the linear $3\alpha$ configuration is energetically unfavored, 
the $\alpha$ distribution is somewhat suppressed in $\theta\sim 0$ and $180^\circ$ regions
at $R=4$, 5 fm. At $R\ge 6$ fm, the angular distribution is  
almost flat in all $\theta$ regions, indicating that the $\alpha$ cluster moves 
freely in angular motion with no angular correlation.

Both $^{11}$B($3/2^-_3$) and $^{12}$C($0^+_2$) contain
significant cluster distribution in the $R=4\text{--}5$ fm region. In this region, the cluster 
around the $2\alpha$ moves almost freely in angular motion with weak angular correlation.
The present result indicates that these two states,
$^{11}$B($3/2^-_3$) and $^{12}$C($0^+_2$), 
have a characteristic of a cluster gas in angular motion,
if we tolerate the exclusion of the $\theta\sim 0$ and $180^\circ$ regions. 

\begin{figure}[tb]
\begin{center}	
	\includegraphics[width=8.5cm]{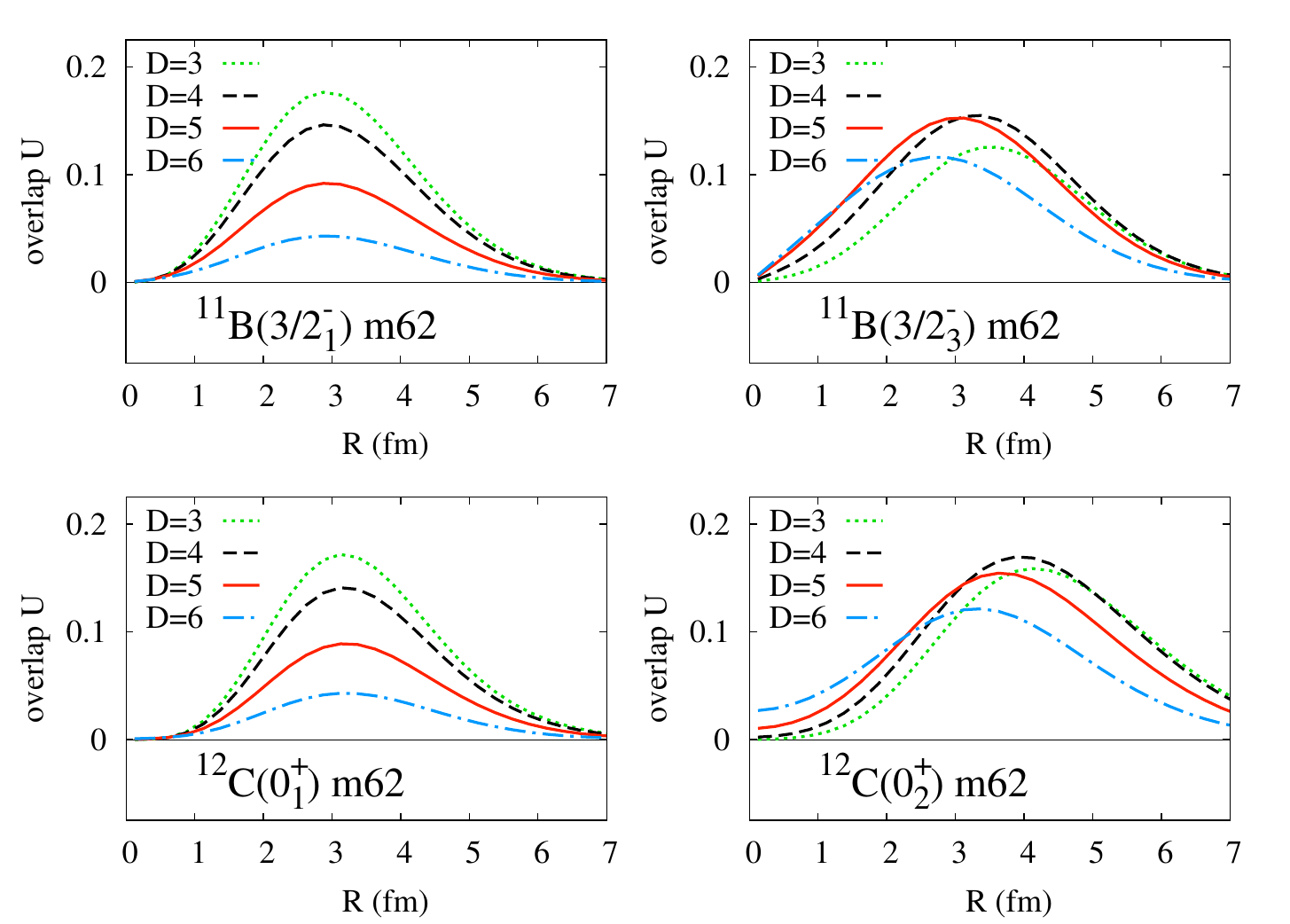}
\end{center}
\vspace{0.5cm}
  \caption{(Color online) $R$ dependence of $U(D,\bvec{R})$ for $^{11}$B($3/2^-_1$),  $^{11}$B($3/2^-_3$), 
$^{12}$C($0^+_1$), and $^{12}$C($0^+_2$) calculated with the m62 interaction.
\label{fig:radial}}
\end{figure}

Another characteristic of a cluster gas is a wide distribution of clusters in 
radial motion. The radial motion of the $t$ cluster around the $2\alpha$ at $\theta=90^\circ$ 
in $^{11}$B($3/2^-_1$) and  $^{11}$B($3/2^-_3$) and that of the $\alpha$ cluster in 
$^{12}$C($0^+_1$) and $^{12}$C($0^+_2$) are shown in Fig.~\ref{fig:radial}.
The $t$ distribution in $^{11}$B($3/2^-_1$) is localized around $R=3$ fm and rapidly 
damps as $R$ increases. The suppression in the small $R$ region comes from the antisymmetrization 
effect from the $2\alpha$. The $\alpha$ distribution in $^{12}$C($0^+_1$) has 
quite similar behavior to $^{11}$B($3/2^-_1$).
In $^{11}$B($3/2^-_3$) and $^{12}$C($0^+_2$), 
the $t$ ($\alpha$) cluster around the $2\alpha$ is distributed  
in the outer region widely, compared to distribution in the ground states. The cluster distribution is
suppressed in the small $R$ region, in particular, in the compact 2$\alpha$ core case with a small $D$
because of the orthogonality to the ground state as well as the antisymmetrization effect. 
The $t$ distribution 
in $^{11}$B($3/2^-_3$) and the $\alpha$ distribution in $^{12}$C($0^+_2$) are outspread widely.
In particular, the radial extent of the $\alpha$ distribution in $^{12}$C($0^+_2$) is remarkable, and 
that of the $t$ distribution in $^{11}$B($3/2^-_3$) is relatively small.
The relatively small extent in $^{11}$B($3/2^-_3$) originates in 
the deeper $t$-$\alpha$ binding than the $\alpha$-$\alpha$ binding 
because of the weaker Pauli blocking effect between $t$ and $\alpha$ clusters.
As a result, the radial spreading of the $t$ cluster in $^{11}$B($3/2^-_3$)
is less remarkable than that of $\alpha$ cluster $^{12}$C($0^+_2$). 

The $D$ dependence of the overlap $U(D,\bvec{R})$
shows the $\alpha$-$\alpha$ radial 
motion in the $2\alpha$ core. As seen in Fig.~\ref{fig:radial}, 
in the ground states of 
$^{11}$B and $^{12}$C,  
the peak hight of the overlap rapidly decrease as $D$ increases,
indicating that two $\alpha$ clusters 
are tightly bound to form a compact $2\alpha$ core.
In contract, 
$^{11}$B($3/2^-_3$) and $^{12}$C($0^+_2$) contain a significant cluster component
for $D=3\text{--}6$ fm, indicating that the $2\alpha$ core in these states 
is a weakly bound $2\alpha$ cluster. The radial extent of the $2\alpha$ 
in $^{11}$B($3/2^-_3$) is as large as that in $^{12}$C($0^+_2$). 
In other words, two $\alpha$ clusters in $^{11}$B($3/2^-_3$) behave as 
an $\alpha$ cluster gas. 

\subsection{Analogies of and differences in $2\alpha+t$ and $3\alpha$ cluster gas states}

As mentioned previously, 
$^{11}$B($3/2^-_3$) and $^{12}$C($0^+_2$) contain a
significant cluster distribution in the $R=4\text{--}5$ fm region. In this region, the cluster 
around the $2\alpha$ core moves rather freely in the angular mode in a wide $\theta$ region,
except for the $\theta\sim 0$ and $180^\circ$ regions.
For $^{11}$B($3/2^-_3$), the $L^{(t)}_z=+1$ motion of the $t$ cluster
in the $(J^\pi,K)=(3/2^-,+3/2)$ component of the $2\alpha+t$ wave function
elementarily has the node structure at the $z$-axis. 
We know that the topological structure of the $P$ wave motion for the $t$ cluster
is different from that of the $S$ wave motion for the $\alpha$ cluster, 
experiencing neither node nor phase change in angular motion.
However, when we consider the 2D cluster motion in the intrinsic frame, 
the angular motion of the $t$ cluster in the $2\alpha+t$ system can be associated 
with that of the $\alpha$ cluster in the $3\alpha$ system, as follows.

The $\alpha$ cluster distribution around the $2\alpha$ core tends to be 
suppressed at $\theta=0$ and $180^\circ$ 
because of the Pauli blocking effect from the $2\alpha$ core on the $z$-axis 
and also because of the energy loss in the linear $3\alpha$ configuration.
Let us consider the angular motion in the extreme case of 
$D=2R$. In the angular distribution of the $\alpha$ cluster moving around the $2\alpha$, 
the probability completely vanishes at $\theta=0$ and $180^\circ$ because of 
the Pauli exclusion principle between nucleons in different $\alpha$ clusters. 
This situation is analogous to the angular distribution of the $t$ cluster in the $P$ wave 
with $L^{(t)}_z=+1$, in which the $t$ probability vanishes at $\theta=0$ and $180^\circ$. 
It should be noted that 
the sign of the $\alpha$ wave function 
is the same in $0< \theta<180^\circ$ and $180^\circ < \theta < 360^\circ$, but 
the sign of the $t$ wave function in $180^\circ < \theta < 360^\circ$ is opposite 
to that in $0< \theta<180^\circ$.
The opposite sign for a $t$ wave function is a consequence of the negative parity 
of the total $2\alpha+t$ system, which originally comes from the negative parity of the lowest allowed 
orbit for the $t$-$\alpha$ intercluster motion,
because the $t$ cluster consists of odd-numbered fermions.
By taking the absolute value of the  $t$-cluster angular wave function,
having two nodes at $\theta=0$ and $180^\circ$ around the $2\alpha$ in the $2\alpha+t$ system, we 
can obtain an angular distribution similar to the $\alpha$ distribution around the $2\alpha$ in the $3\alpha$ system, which 
means that the one-dimensional $t$ motion in the angular mode 
around the fixed $2\alpha$ core can be mapped on the $\alpha$ motion around $2\alpha$.
In three-dimensional space,
the phase of the $t$ wave function of the $L^{(t)}_z=+1$ orbit changes by $2\pi$
in the rotation around the $z$-axis. 
Again by taking the absolute value of the $t$ wave function, 
it may be possible to consider a mapping of the $t$-cluster motion onto the $\alpha$-cluster motion, if the $\alpha$ probability is suppressed at $\theta=0$ and $180^\circ$. 

In the $3\alpha$ system for $^{12}$C($0^+_2$), the $\alpha$ probability
is somewhat suppressed at $\theta=0$ and $180^\circ$ in the $R\le 5$ fm region. 
In this region, the angular motion of the $t$ cluster in $^{11}$B($3/2^-_3$) 
has some association with the $\alpha$-cluster motion in $^{12}$C($0^+_2$), as described 
previously. That is, the cluster motion in both cases 
is characterized by the wide angular distribution,
except for the $\theta=0$ and $180^\circ$ regions.
However, 
this $t$-cluster and $\alpha$-cluster association in the angular motion around 
the $2\alpha$ breaks down in the asymptotic region, far from the $2\alpha$ core. 
The $\alpha$ cluster distribution is isotropic and 
goes to the ideal $S$-wave motion in the asymptotic region,
where the blocking effect from the core vanishes; by comparison, 
the $t$ cluster in a $P$ wave always has the
 node structure in the angular mode, even 
in the asymptotic region. 
Nevertheless, 
the $t$ cluster in $^{11}$B($3/2^-_3)$ is distributed in a wide $\theta$ region around the $2\alpha$, indicating that the total system has a non-geometric cluster structure
with weak angular correlation rather than a geometric structure.
If we tolerate the exclusion of $\theta\sim 0$ and $180^\circ$ regions and extend the 
concept of the cluster gas as a developed cluster structure 
with weak angular correlation,  
$^{11}$B($3/2^-_3$) can be interpreted as a kind of cluster gas of $2\alpha+t$.

Therefore we can state the following: 
\begin{itemize}
\item $^{11}$B($3/2^-_3$) and $^{12}$C($0^+_2$) are $2\alpha+t$ and $3\alpha$ 
cluster states with weak angular correlation, like a cluster gas.
\end{itemize}

In addition to the similarity of and difference in 
the angular correlation mentioned previously, the following similarities and differences
exist in the $2\alpha+t$ cluster structure of $^{11}$B($3/2^-_3$) and the $3\alpha$ 
cluster structure of $^{12}$C($0^+_2$). 

\begin{itemize}
\item
In the radial motion of the $t$ cluster around the $2\alpha$, the $t$ distribution is 
outspread widely in $^{11}$B($3/2^-_3$), compared with the ground state, and it 
is similar to the radial motion of the $\alpha$ cluster in $^{12}$C($0^+_2$). 
However, 
the radial extent of the $t$ distribution in $^{11}$B($3/2^-_3$) is relatively smaller 
than that of the $\alpha$ distribution in $^{12}$C$(0^+_2$) because the 
$t$-$\alpha$ binding is deeper than the $\alpha$-$\alpha$ binding, which occurs 
because of the weaker Pauli blocking between $t$ and $\alpha$ clusters. 

\item The $D$ dependence of the overlap $U(D,\bvec{R})$ indicates that 
two $\alpha$ clusters in $^{11}$B are bound as weakly as those in $^{12}$C. 

\end{itemize}

In conclusion, $^{11}$B($3/2^-_3$) is interpreted as a three-body cluster gas of $2\alpha+t$
in weak angular correlation and the radial extent of clusters.
The cluster gas feature is more prominent in $^{12}$C($0^+_2$) than in $^{11}$B($3/2^-_3$).

\section{Summary}\label{sec:summary}

In this study, we reanalyzed
cluster features of $3/2^-$ states of $^{11}$B and  $0^+$ states of $^{12}$C,
obtained with the AMD+VAP method in 
previous study. The $2\alpha+t$ cluster structures of $^{11}$B
were compared with the $3\alpha$ cluster structures of $^{12}$C in the analysis of 
cluster distribution.
We were particularly attentive to the $t$-cluster motion around 
the $2\alpha$ core in the $2\alpha+t$ system. 
We considered the 2D cluster motion in the intrinsic frame, 
and investigated the cluster distribution on the 2D plane. 
To discuss the angular motion and radial motion of the $t$ cluster in $^{11}$B
and those of the $\alpha$ cluster in $^{12}$C, 
we studied the dependencies of the cluster distribution 
on the distance $R$ and the angle $\theta$ for the cluster 
position from the $2\alpha$.
In particular, we discussed the $\theta$ dependence 
to clarify the angular motion of the $t$ cluster around the $2\alpha$. 

$^{11}$B($3/2^-_3$) and $^{12}$C($0^+_2$) contain 
significant $t$- and $\alpha$-cluster distributions 
in the radial distance $R=4\text{--}5$ fm region. 
In this region, the $t$-cluster motion in $^{11}$B($3/2^-_3$) 
is characterized by the wide angular distribution,
except for the $\theta=0$ and $180^\circ$ regions. It is associated 
with the angular motion of the $\alpha$ cluster in $^{12}$C($0^+_2$), where the
$\alpha$ distribution in the $\theta=0$ and $180^\circ$ regions is somewhat
suppressed. By comparing the $2\alpha+t$ cluster structure in $^{11}$B($3/2^-_3$)
and the $3\alpha$ cluster structure in $^{12}$C($0^+_2$),
the following similarities and differences are found. 
\begin{itemize}
\item In angular motion, 
$^{11}$B($3/2^-_3$) and $^{12}$C($0^+_2$) are $2\alpha+t$ and $3\alpha$ 
cluster states with weak angular correlation, like a cluster gas.
\item
In radial motion of the $t$ cluster around the $2\alpha$, the $t$ distribution is 
outspread widely in $^{11}$B($3/2^-_3$), compared with the ground state. However, 
the radial extent of the $t$ distribution in $^{11}$B($3/2^-$) is less than 
that of the $\alpha$ distribution in $^{12}$C$(0^+_2$) because the
$t$-$\alpha$ binding is deeper than the $\alpha$-$\alpha$ binding. 
\item The $D$ dependence of the overlap $U(D,\bvec{R})$ indicates that 
two $\alpha$ clusters in $^{11}$B are bound weakly as those in $^{12}$C. 
\end{itemize}

In conclusion, $^{11}$B($3/2^-_3$) is interpreted as a three-body cluster gas of $2\alpha+t$
in sense of weak angular correlation and radial extent of the clusters.
The cluster gas feature is more prominent in $^{12}$C($0^+_2$) than in $^{11}$B($3/2^-_3$).

\section*{Acknowledgments} 
The authors would like to thank Prof.~Schuck for fruitful discussions.
The computational calculations of this work were performed using the
supercomputers at YITP.
This work was supported by 
JSPS KAKENHI Grant Numbers 25887049 and 26400270.


\begin{thebibliography}{0}
\bibitem{Fujiwara80}
Y. Fujiwara {\em et al.}, Prog. Theor. Phys. Suppl. {\bf 68}, 29 (1980).

\bibitem{Freer:2014qoa} 
  M.~Freer and H.~O.~U.~Fynbo,
  Prog.\ Part.\ Nucl.\ Phys.\  {\bf 78}, 1 (2014).
  
\bibitem{morinaga56}
H. Morinaga, Phys. Rev. {\bf 101}, 254 (1956).
\bibitem{morinaga66}
H. Morinaga, Phys. Lett. {\bf 21}, 78 (1966).
\bibitem{suzuki72}
Y. Suzuki, H. Horiuchi and K. Ikeda, Prog. Theor. Phys. {\bf 47}, 
1517 (1972).   

\bibitem{kamimura-RGM1}
Y. Fukushima and M. Kamimura,
{\it Proc. Int. Conf. on Nuclear Structure, Tokyo, 1977,
edited by T. Marumori} J. Phys. Soc. Jpn. {\bf 44}, 225 (1978).
\bibitem{kamimura-RGM2}
M. Kamimura, Nucl. Phys. {\bf A351}, 456 (1981).
\bibitem{uegaki1}
E. Uegaki, S. Okabe, Y. Abe and H. Tanaka, Prog. Theor. Phys. {\bf 57},
1262 (1977).
\bibitem{uegaki2}
E. Uegaki, Y. Abe, S. Okabe and H. Tanaka, Prog. Theor. Phys. {\bf 59},
 1031 (1978).
\bibitem{uegaki3}
E. Uegaki, Y. Abe, S. Okabe and H. Tanaka, Prog. Theor. Phys. {\bf 62}, 1621 (1979).

\bibitem{descouvemont87}
P. Descouvemont and D. Baye, Phys. Rev. C {\bf 36}, 54 (1987).


\bibitem{kurokawa04}
C. Kurokawa and K. Kato,
Nucl. Phys. {\bf A738}, 455 (2004).

\bibitem{kurokawa05}
C. Kurokawa and K. Kato,
Phys. Rev. C {\bf 71}, 021301(R) (2005).

\bibitem{Arai:2006bt} 
  K.~Arai,
  Phys.\ Rev.\ C {\bf 74}, 064311 (2006).

\bibitem{Enyo-c12}
 Y. Kanada-En'yo,
Phys. Rev. Lett. {\bf 81}, 5291 (1998).
\bibitem{KanadaEn'yo:2006ze} 
  Y.~Kanada-En'yo,
  Prog.\ Theor.\ Phys.\  {\bf 117}, 655 (2007)
  [Erratum-ibid.\  {\bf 121}, 895 (2009)].

\bibitem{Neff-c12}
T. Neff and H. Feldmeier, Nucl. Phys. {\bf A738}, 357 (2004).

\bibitem{Chernykh:2007zz} 
  M.~Chernykh, H.~Feldmeier, T.~Neff, P.~von Neumann-Cosel and A.~Richter,
  Phys.\ Rev.\ Lett.\  {\bf 98}, 032501 (2007).

\bibitem{Tohsaki01}
A. Tohsaki, H. Horiuchi, P. Schuck, and G. R\"opke, 
Phys. Rev. Lett. {\bf 87}, 192501 (2001).

\bibitem{Funaki:2002fn} 
  Y.~Funaki, H.~Horiuchi, A.~Tohsaki, P.~Schuck and G.~Ropke,
  Prog.\ Theor.\ Phys.\  {\bf 108}, 297 (2002).

\bibitem{Yamada:2003cz} 
  T.~Yamada and P.~Schuck,
  Phys.\ Rev.\ C {\bf 69}, 024309 (2004).

\bibitem{Funaki:2003af} 
  Y.~Funaki, A.~Tohsaki, H.~Horiuchi, P.~Schuck and G.~Ropke,
  Phys.\ Rev.\ C {\bf 67}, 051306 (2003).
\bibitem{Funaki:2009zz} 
  Y.~Funaki, H.~Horiuchi, W.~von Oertzen, G.~Ropke, P.~Schuck, A.~Tohsaki and T.~Yamada,
  Phys.\ Rev.\ C {\bf 80}, 064326 (2009).
\bibitem{Yamada:2011bi} 
  T.~Yamada, Y.~Funaki, H.~Horiuchi, G.~Roepke, P.~Schuck and A.~Tohsaki,
  Lect.\ Notes Phys.\  {\bf 848}, 229 (2012).
  

\bibitem{Ropke98}
G. R\"opke, A. Schnell, P. Schuck, and P. Nozieres, Phys. Rev. Lett. {\bf 80},
3177 (1998).

\bibitem{Kawabata:2005ta} 
  T.~Kawabata, H.~Akimune, H.~Fujita, Y.~Fujita, M.~Fujiwara, K.~Hara, K.~Hatanaka and M.~Itoh {\it et al.},
  Phys.\ Lett.\ B {\bf 646}, 6 (2007).

\bibitem{KanadaEn'yo:2006bd} 
  Y.~Kanada-En'yo,
  Phys.\ Rev.\ C {\bf 75}, 024302 (2007).

\bibitem{KanadaEn'yo:2007ie} 
  Y.~Kanada-En'yo,
  Phys.\ Rev.\ C {\bf 76}, 044323 (2007).

\bibitem{Itagaki:2007rv} 
  N.~Itagaki, M.~Kimura, C.~Kurokawa, M.~Ito and W.~von Oertzen,
  Phys.\ Rev.\ C {\bf 75}, 037303 (2007).

\bibitem{Wakasa:2007zza} 
  T.~Wakasa, E.~Ihara, K.~Fujita, Y.~Funaki, K.~Hatanaka, H.~Horiuchi, M.~Itoh and J.~Kamiya {\it et al.},
  Phys.\ Lett.\ B {\bf 653}, 173 (2007).


 \bibitem{Funaki:2008gb} 
  Y.~Funaki, T.~Yamada, H.~Horiuchi, G.~Ropke, P.~Schuck and A.~Tohsaki,
  Phys.\ Rev.\ Lett.\  {\bf 101}, 082502 (2008).

\bibitem{Funaki:2010px} 
  Y.~Funaki, T.~Yamada, A.~Tohsaki, H.~Horiuchi, G.~Ropke and P.~Schuck,
  Phys.\ Rev.\ C {\bf 82}, 024312 (2010).
  
\bibitem{Yamada:2010vc} 
  T.~Yamada and Y.~Funaki,
  Phys.\ Rev.\ C {\bf 82}, 064315 (2010).

\bibitem{Ohkubo:2010zz} 
  S.~Ohkubo, Y.~Hirabayashi and Y.~Hirabayashi,
  Phys.\ Lett.\ B {\bf 684}, 127 (2010).
  
\bibitem{Suhara:2012zr} 
  T.~Suhara and Y.~Kanada-En'yo,
  Phys.\ Rev.\ C {\bf 85}, 054320 (2012).

\bibitem{Ichikawa:2012mk} 
  T.~Ichikawa, N.~Itagaki, Y.~Kanada-En'yo, T.~.Kokalova and W.~von Oertzen,
  Phys.\ Rev.\ C {\bf 86}, 031303 (2012).

\bibitem{Kobayashi:2012di} 
  F.~Kobayashi and Y.~Kanada-En'yo,
  Phys.\ Rev.\ C {\bf 86}, 064303 (2012).

\bibitem{Kobayashi:2013iwa} 
  F.~Kobayashi and Y.~Kanada-En'yo,
  Phys.\ Rev.\ C {\bf 88}, no. 3, 034321 (2013).

\bibitem{nishioka79}
H. Nishioka, S. Saito and M. Yasuno, Prog. Theor. Phys. {\bf 62} 424 (1979).

\bibitem{Kawabata:2004pc} 
  T.~Kawabata, H.~Akimune, H.~Fujimura, H.~Fujita, Y.~Fujita, M.~Fujiwara, K.~Hara and K.~Y.~Hara {\it et al.},
  Phys.\ Rev.\ C {\bf 70}, 034318 (2004).
\bibitem{Fujita04}
Y. Fujita et al., Phys. Rev. C {\bf 70}, 011306(R) (2004).

\bibitem{Brink66}
                D.M. Brink, {\em Proceedings of the International School of
                Physics ``Enrico Fermi''}, Varenna, 1965, Course 36,
                Ed. by C.Bloch (Academic Press, New York 1966)



\bibitem{KanadaEnyo:1995tb} 
  Y.~Kanada-Enyo, H.~Horiuchi and A.~Ono,
  Phys.\ Rev.\ C {\bf 52}, 628 (1995).

\bibitem{KanadaEnyo:1995ir} 
  Y.~Kanada-Enyo and H.~Horiuchi,
  Phys.\ Rev.\ C {\bf 52}, 647 (1995).

\bibitem{ENYOsup}
Y. Kanada-En'yo and  H. Horiuchi, Prog. Theor. Phys. Suppl.{\bf 142},
 205 (2001).
\bibitem{AMDrev}
Y. Kanada-En'yo, M. Kimura and H. Horiuchi, Comptes rendus Physique Vol.4, 
497 (2003).

\bibitem{KanadaEn'yo:2012bj}
  Y.~Kanada-En'yo, M.~Kimura and A.~Ono,
  PTEP {\bf 2012} (2012) 01A202.

\bibitem{Suhara-new}
in preparation.

\bibitem{MVOLKOV} T. Ando, K. Ikeda and A. Tohsaki,
	Prog. Theory. Phys. {\bf 64}, 1608   (1980).


\bibitem{LS}
 N. Yamaguchi, T. Kasahara, S. Nagata and Y. Akaishi,
 {Prog. Theor. Phys.} {\bf 62}, 1018  (1979);
 R. Tamagaki, {Prog. Theor. Phys.} {\bf 39}, 91  (1968).




 \bibitem{nucl-data-charge-radius}
I.~Angeli, Atomic Data and Nuclear Data Tables {\bf 87}, 185 (2004).





\end{thebibliography}
\end{document}